\documentclass[letterpaper,11pt]{article}
\pdfoutput=1

\usepackage{jheppub}
\usepackage{multirow}

\usepackage{subcaption}
\usepackage[countmax]{subfloat}
\usepackage{amssymb}
\usepackage{amsmath}
\usepackage{color}
\usepackage{graphicx}
\usepackage{verbatim}
\usepackage{amsthm}
\usepackage{slashed}
\usepackage{hyperref}

\def\cI{\mathcal{I}}

\def\cO{\mathcal{O}}

\def\nn{{\nonumber}}
\def\mcdot{\!\cdot\!}

\newcommand{\df}{\mathrm{d}}

\newcommand{\e}{\epsilon}

\newcommand{\zero}{{(0)}}
\newcommand{\one}{{(1)}}

\newcommand{\as}{\alpha_{\scriptscriptstyle S}}

\def\dd{\mathrm{d}}

\preprint{}

\title{
Massive charged-current coefficient functions in deep-inelastic scattering 
at NNLO and impact
on strange-quark distributions
}

\author{Jun Gao}

\affiliation{
INPAC, Shanghai Key Laboratory for Particle Physics and Cosmology, School of
Physics and Astronomy, Shanghai Jiao-Tong University, Shanghai 200240, China
}

\emailAdd{jung49@sjtu.edu.cn}

\abstract{
We present details on calculation of next-to-next-to-leading order QCD corrections to
massive charged-current coefficient functions in deep-inelastic scattering. 
Especially we focus on the application to charm-quark production in neutrino scattering on
fixed target that can be measured via the dimuon final state.
We construct a fast interface to the calculation so for any parton
distributions the cross sections can be evaluated within milliseconds
by using the pre-generated interpolation grids.
We discuss agreements of various theoretical predictions
with the NuTeV and CCFR dimuon data and the impact of the results
on determination of the strange-quark distributions.  
}
  

\begin{document} 

\maketitle

\section{Introduction}
Deep-inelastic scattering (DIS) remains the most important probe of
quantum chromodynamics (QCD) since it was pioneered almost fifty years ago.
Various dedicated experiments have been carried out since then aiming to
study the internal structure of nucleons, e.g., the HERA experiments
by colliding electron or positron with proton.
The HERA measurements on neutral-current (NC) and charged-current (CC) DIS
provide the backbone constraint in modern determination of the parton distribution
functions (PDFs)~\cite{1506.06042,Gao:2017yyd}.
The heavy quark, especially charm quark plays an important role in
describing the structure functions of proton measured in DIS within
the framework of QCD factorization.
In perturbative QCD calculations heavy-quark mass dependence of the
DIS coefficient functions are crucial for analyses of the DIS data,
in the inclusive structure function measurements and even more in
the open production of heavy quarks, and ensure a precise determination
of PDFs that is vital for the ongoing programs at the Large
Hadron Collider (LHC).
In the neutral-current case the DIS coefficient functions are
known up to ${\mathcal O}(\as^2)$ with exact heavy-quark mass
effects~\cite{Laenen:1992zk,Laenen:1992xs}.
For the charged-current case the heavy-quark mass effects have
been calculated to ${\mathcal O}(\as)$ in
Refs.~\cite{Gottschalk:1980rv,Gluck:1997sj,Blumlein:2011zu},
to approximate ${\mathcal O}(\as^2)$ in Ref.~\cite{Alekhin:2014sya}.
Recently the exact ${\mathcal O}(\as^2)$ results with full mass
dependence have been completed~\cite{1601.05430}.
The $\mathcal {O} (\alpha_s^3)$
results are also available for structure function
$xF_3$ at large momentum transfer~\cite{Behring:2015roa}.

Among all DIS experiments there exist one specific measurement,
charm-quark production in DIS of a neutrino from a heavy-nucleus.
It provides direct access to the strange quark content of the nucleon
which is poorly constrained by the inclusive structure function data.
At lowest order, the relevant partonic process is neutrino interaction with
a strange quark, $\nu s \rightarrow c X$, mediated by the weak charged current.
Experimentally one can require a semi-leptonic decay of the charm quark to muon
that gives the so-called {\it dimuon} final state as measured by CCFR~\cite{Goncharov:2001qe},
NuTeV~\cite{Mason:2006qa}, CHORUS~\cite{KayisTopaksu:2008aa}, and NOMAD~\cite{Samoylov:2013xoa}
collaborations.
In global determination of PDFs it is from those dimuon data which prefers
a suppressed strange-quark distribution than $u$ and $d$ sea-quarks
~\cite{Dulat:2015mca,Harland-Lang:2014zoa,Ball:2014uwa}.
That agrees with predictions from various models suggesting that the
strange PDFs are suppressed compared to those of light sea quarks due
to its larger mass~\cite{Carvalho:1999he,Vogt:2000sk,Chen:2009xy}.
The strange quark PDFs can play an important role in LHC phenomenology,
contributing, for example, to the total PDF uncertainty 
in $W$ or $Z$ boson production~\cite{Nadolsky:2008zw,1203.1290}, 
and to systematic uncertainties in 
precise measurements of the $W$ boson mass and weak-mixing
angle~\cite{Krasny:2010vd,Bozzi:2011ww,Baak:2013fwa}.

On another hand, thanks to the LHC we can also extract the strange
quark PDFs independently from collider data only, e.g, via a combined
analysis of HERA DIS data and the $W$, $Z/\gamma^*$ boson production data
from the LHC.
The latter can provide constraints on the strange quark PDFs due to
its high precision and the fact that differential distributions can
separate different sea flavors.
The ATLAS collaboration have reported such a study, ATLAS-epWZ16~\cite{1612.03016},
using the HERA I and II combined data and the updated 7 TeV measurements on
$W$ and $Z/\gamma^*$ differential cross sections.
Interestingly, an unsuppressed strange quark PDF is preferred
with $R_s\equiv (s+\bar s)/(\bar u+\bar d)$ measured to
be $1.13^{+0.08}_{-0.13}$ at $x=0.023$ and $Q^2=1.9\,{\rm GeV}^2$.
Similar conclusion has been reached in an earlier ATLAS study
based on a smaller sample of $W$ and $Z$ data~\cite{1203.4051}.
In comparison the values of $R_s$ from global PDF determination
are $0.55\pm 0.21$, $0.57\pm 0.17$, $0.60\pm 0.13$, and $0.63\pm 0.03$
for CT14~\cite{Dulat:2015mca}, MMHT2014~\cite{Harland-Lang:2014zoa},
NNPDF3.1~\cite{Ball:2017nwa}, and ABMP16~\cite{1701.05838}.
In NNPDF3.1 they also performed an alternative fit using exactly
the same data sets as in the ATLAS-epWZ16 analysis.
They confirmed the pull on the central values of the strange-quark PDFs
by the ATLAS data but arrived at a much larger uncertainties.
The discrepancies seen between the determinations of strangeness
from fixed-target dimuon data and the ATLAS data have attracted a
lot of attentions recently.
It was suggested in Ref.~\cite{1708.01067} that the ATLAS determination
may be biased by the special parametrization form of PDFs adopted.
While future LHC data might be helpful to clarify whether there are
indeed tensions between those two determinations, it is also
important to investigate various theoretical uncertainties
especially in the case of charm quark production in DIS.   

In Ref.~\cite{1601.05430} we have reported a first application of our
${\mathcal O}(\as^2)$ results on the massive coefficient
functions of charged-current DIS.
We calculated the next-to-next-to-leading-order (NNLO)
QCD corrections to charm-quark production in DIS of a 
neutrino from a nucleon.
The calculation is based on a phase-space slicing method and 
fully-differential Monte Carlo integration.
We found the NNLO corrections can change the cross sections
by up to 10\% depending on the kinematic region considered.
In this paper we provide further elaboration of the methods
and numerical results of our NNLO calculation.
Moreover, we implement our calculation into a fast interface
based on grid interpolation that ensures repetition of the
calculation within millisecond for arbitrary PDFs.
It allows a first study of effects of the NNLO massive
coefficient functions on determination of the strange-quark
PDFs in the context of Hessian profiling, and can be
used in future global analysis of PDFs. 

In the remaining paragraphs we outline the method used in the
calculation, present detailed numerical results on QCD corrections
on cross sections of charm-quark production in charged-current DIS,
demonstrate the accuracy of the grid interpolation,
study the agreements of data and theory with various PDFs, and
finally study effects of the NNLO corrections on extraction of the
strange-quark PDFs.

\section{NNLO calculation}\label{sec:cal}
We have presented briefly the framework of our NNLO calculation
together with selected numerical results in Ref.~\cite{1601.05430}.
Here we give more details on the theoretical ingredients of the
calculation as well as more numerical results focusing on kinematic region
of the fixed-target dimuon measurements.
We also discuss the applicable kinematic range of our calculation
utilizing fixed-flavor number scheme for heavy quarks and the
possible improvement by extending to a variable-flavor number scheme.   

\subsection{Theoretical framework}
\label{sec:qcd-corr-single}

The perturbative calculation utilizes a generalization of the phase-space
slicing method to NNLO as motivated by the $q_T$ subtraction method proposed
in~\cite{Catani:2007vq}.
There have been quite a few recent applications of similar methods on
either decay processes~\cite{1210.2808},
scattering at lepton colliders~\cite{1408.5150,1410.3165},
and hadron colliders~\cite{1504.02131,1505.04794,1606.08463,1607.06382,1708.09405,Li:2017lbf}.
The key ingredient of above methods is the use of soft-collinear effective
theory~(SCET) and heavy-quark effective theory~(HQET)~\cite{hep-ph/0005275,hep-ph/0011336,hep-ph/0109045,hep-ph/0202088}
to systematically factorize the cross section and derive its perturbative expansion in
fully unresolved region of QCD radiations as was proposed in Ref.~\cite{1210.2808}.
Note here we adopt HQET for purpose of extracting the soft singularities
in the perturbative expansion which is irrelevant to the actual
mass of the charm quark.
Other approaches on handling singularities in the fully unresolved region
include sector-improved FKS subtraction method~\cite{1005.0274,1111.7041}, 
sector decomposition method~\cite{hep-ph/0402265,hep-ph/0311311}, antenna subtraction
method~\cite{hep-ph/0505111,1301.4693}, colorful subtraction method~\cite{1501.07226}, and Projection-to-Born
method~\cite{1506.02660}.
   
\begin{figure}[ht]
  \centering
  \includegraphics[width=0.3\textwidth]{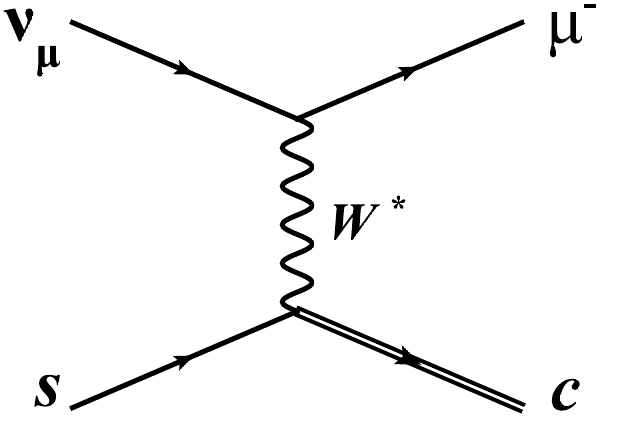}
  \caption{LO diagram for charm quark production through weak charged current
  in DIS. The thick solid line denotes the charm quark.
   \label{fig:5}}
  \end{figure}

Charm quark production at leading order (LO) through weak charged current in DIS can be
represented by the diagram in Fig.~\ref{fig:5}.
There are also Cabibbo suppressed contributions from $d$ quark initial
state.
The production of charm anti-quark is similar. 
In our phase-space slicing method we first define a resolution variable
which can isolate the unresolved phase space. 
As was discussed in Ref.~\cite{1601.05430}, the appropriate resolution
variable in this case is a fully inclusive version of beam 
thrust~\cite{Stewart:2009yx} or N-jettiness~\cite{1004.2489}, 
\begin{align}
  \label{eq:16}
  \tau =  \frac{2 \, p_X \mcdot p_n }{ m^2_c - q^2} ,\qquad \text { with  } 
  p_n  =
\Big(\bar{n} \mcdot (p_c - q)\Big)    \frac{ n^\mu }{2} \,,
\end{align}
which differs from the standard beam thrust or N-jettiness in that no
partition in the phase space of final-state radiation is imposed, as there is
only one collinear direction in the problem. 
In Eq.~\eqref{eq:16}, $p_X$ is the momentum of total QCD radiation in the
final state, $p_c$ is the momentum of the charm quark, and $q$ is the momentum
transfer as carried by a virtual $W$ boson.
$p_n$ is a momentum align with the incoming beam, whose large
lightcone component equals the large lightcone component of the incoming momentum
entering the $Wsc$ vertex. 
Here the lightcone direction $n$ is chosen as the direction of the
incoming beam, and $\bar{n} = (1, -\vec{n})$.

With the definition for $\tau$, the differential cross section for
any infrared-safe observable $O$ can be separated into resolved and
unresolved part,
\begin{align}
  \label{eq:17}
    \frac{\dd\sigma}{\dd O} = & \int^{\tau_{\rm cut}}_0 \dd \tau \, \frac{\dd^2 \sigma}{\dd O
  \, \dd \tau} + \int^{\tau_{\rm max}}_{\tau_{\rm cut}} \dd \tau \, \frac{\dd^2 \sigma}{\dd O
  \, \dd \tau} 
\nn\\
= & \left. \frac{\dd\sigma}{\dd O} \right|_{\rm unres.} +
    \left. \frac{\dd\sigma}{\dd O} \right|_{\rm res.} \,.
\end{align}
Further we can write down a factorization formula for the unresolved
contribution, up to power corrections of
the form $\tau_{\rm cut} \ln^k \tau_{\rm cut}$,
\begin{align}
  \label{eq:18}
  \left. \frac{\dd\sigma}{\dd O} \right|_{\rm unres.}  =  & \int \dd z
  \,  \frac{\dd\sigma^\zero (z)}{\dd O} H(y,\mu) \int^{\tau_{\rm
  cut}}_0 \dd \tau \, \dd t \, \dd k_s   \, B_q( t, z,
  \mu) S(k_s, \mu) 
\nn\\
& \mcdot \delta \left(\tau - \frac{t +2 k_s E_d}{m_c^2 - q^2} \right) + \cO( \tau_{\rm cut} \ln^k \tau_{\rm cut} )  \,,
\end{align}
where $E_d$ is the energy of the $s$ quark entering the $Wcs$ vertex. 
The derivation of this factorization formula is very similar to the
derivation of beam thrust of N-jettiness factorization~\cite{1004.2489}. 
In Eq.~\eqref{eq:18}, $\dd\sigma^\zero(z)/\dd O$ is the Born level partonic
differential cross section for the process
\begin{align}
  \label{eq:19}
 s( zP_N)  +W^*( q )\to c(p_c)  \,,
\end{align}
where $P_N$ is the momentum of the incoming hadron.
The variable $y$
is defined as $y = q^2/m_c^2 < 0$. 
The hard function $H(y,\mu)$ for charm quark
production can be straightforwardly related to the hard function for
bottom quark decay through analytic continuation.
We refer readers to
Refs.~\cite{Bonciani:2008wf,Asatrian:2008uk,Beneke:2008ei,Bell:2008ws}
for the full two-loop results
\footnote{In our calculation, we use the
result of Ref.~\cite{Asatrian:2008uk}, kindly provided to us by Ben
Pecjak in a convenient computer readable form.}.

The soft function $S$ is defined as a vacuum matrix element of Wilson
loops.  
In a practical calculation, they can be obtained by taking the eikonal limit of the
real corrections, with the insertion of a 
measurement function $\delta( k_s - k \mcdot n)$, where $k_s$ is the
total momentum of the soft radiation in the final state.
For instance at one-loop the soft function can be calculated from the diagrams
\begin{align}
  \label{eq:21}
    S^\one(k_s,\mu) =\mu^{2 \e} \int \frac{\dd^{4 - 2 \e} k}{(2\pi)^{4 -  2
  \e}} (2\pi) \Theta(k^0) \delta(k^2) \delta( k_s - k \mcdot n) \left| \parbox[h]{0.18\textwidth}{
\includegraphics[width=0.15\textwidth]{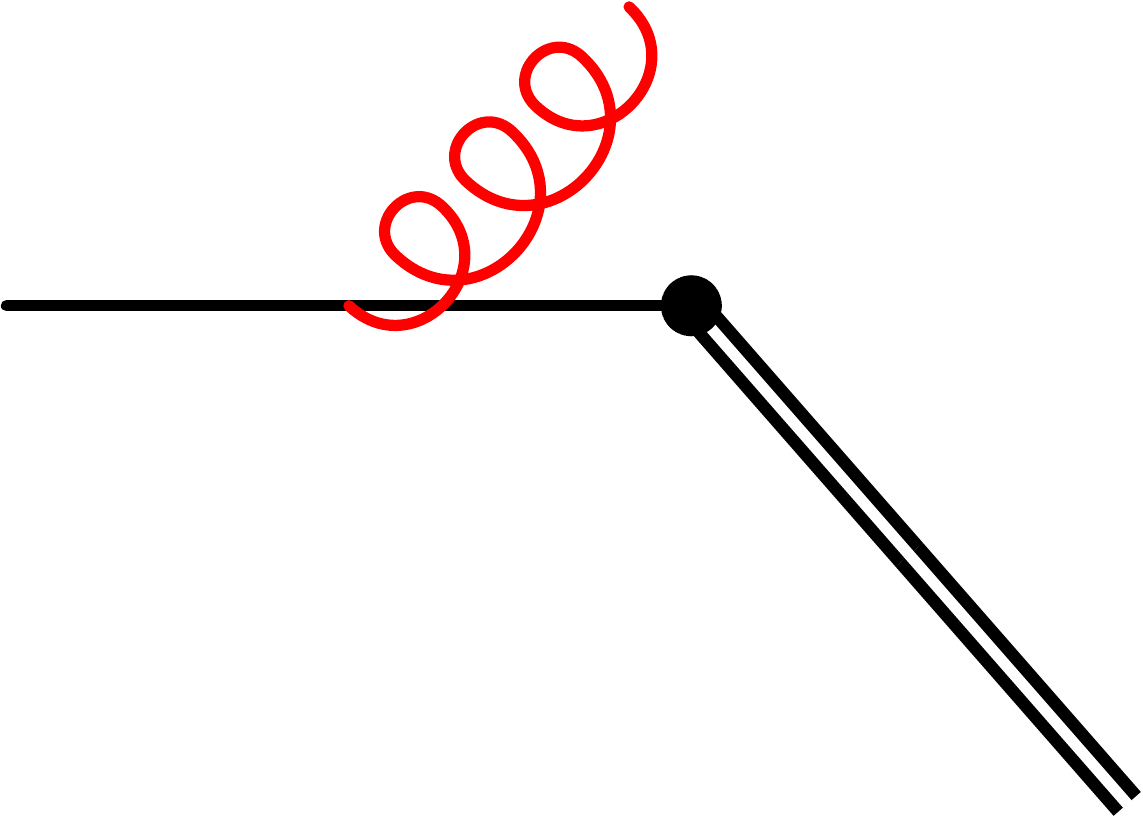} 
}
+ 
\parbox[h]{0.18\textwidth}{
\includegraphics[width=0.15\textwidth]{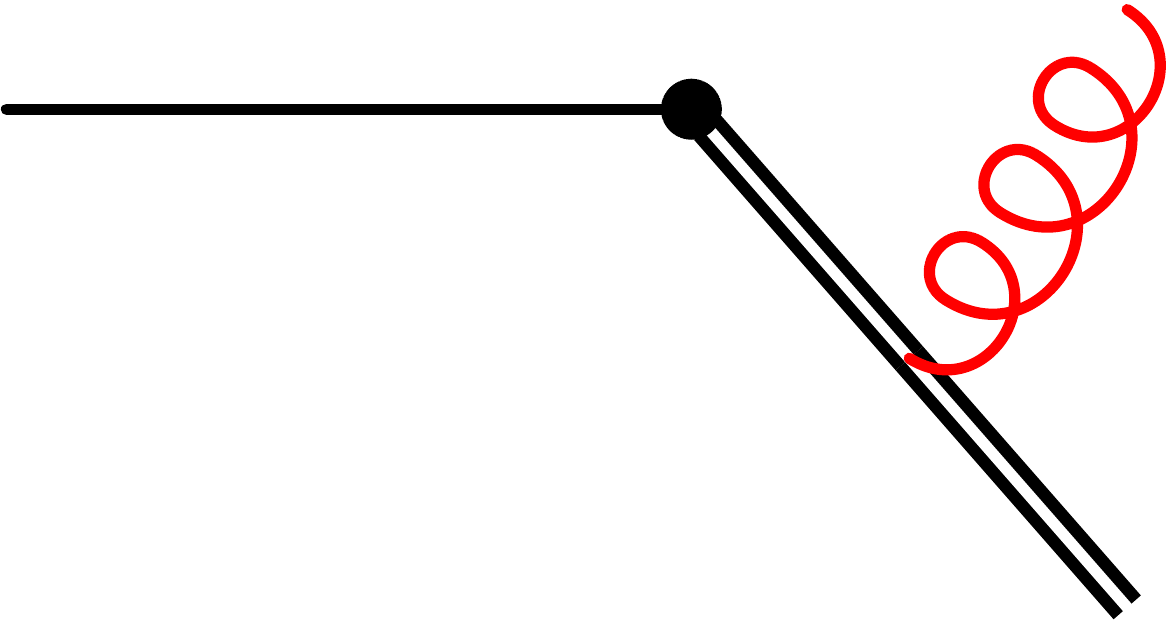} }
\right|^2 \,,
\end{align}
where the lightlike direction $n$ is pointing in the incoming beam
direction. 
We use a double line to denote a timelike Wilson line, and a solid
real line to denote a lightlike Wilson line.
Note that the definition
for the soft function is not Lorentz invariant.  
The violation of
Lorentz invariance comes only from the measurement function
$\delta(k_s - k \mcdot n)$.
However, the full result when combining with the $\delta$ function
in Eq.~(\ref{eq:18}) is Lorentz invariant. 
We quote the result for the soft function through one loop in charm-quark
rest frame as below,
\begin{align}
  \label{eq:11}
  S(k_s,\mu) = \delta(k_s) + \frac{\as}{4\pi} C_F \left( - 8 \left[
  \frac{\ln(k_s/\mu)}{k_s}\right]_\star^{[k_s,\mu]} - 4 \left[
  \frac{1}{k_s}\right]_\star^{[k_s,\mu]} - \frac{\pi^2}{6} \delta(k_s) \right) + \cO(\as^2) \,,
\end{align}
where the star distribution is defined as
\begin{align}
  \label{eq:12}
  \int^\mu_0 \dd k_s \, [f(k_s)]_\star^{[k_s,\mu]} g(k_s)  =   \int^\mu_0 \dd
  k_s \, f(k_s)( g(k_s) - g(0) ) \,.
\end{align}
We refer to Ref.~\cite{Becher:2005pd} for the full two-loop soft
function. 

The beam function $B$ is defined as the matrix element of collinear field
in a hadron state~(proton in our case), with the virtuality $t = 2 p_n
\mcdot l$ of the beam jet measured~\cite{Stewart:2009yx}, where $l$ is the momentum of
final state collinear radiation, and $p_n$ is defined in
Eq.~\eqref{eq:16}. 
The beam function can be written as convolution of
perturbative coefficient functions and the usual PDFs,
\begin{align}
  \label{eq:27}
  B_i(t, x, \mu) = \sum_j \int \frac{\df\xi}{\xi} \, \cI_{ij} \left(t,
  \frac{x}{\xi}, \mu\right) f_j ( \xi, \mu) + \cO \left(
  \frac{\Lambda_{\rm QCD}^2}{t} \right) \,.
\end{align}
For example, the one-loop quark-to-quark coefficient function can be calculated
through the diagrams
\begin{align}
  \label{eq:28}
    \cI_{qq}^\one\left(t, z, \mu\right) = & \int \frac{\dd^{4 - 2 \e} l}{(2\pi)^{4 -  2
  \e}} (2\pi) \Theta(l^0) \delta(l^2) \delta(t - 2 p_n \mcdot l )
  \delta \big( l \mcdot \bar{n}  - (1-z) p_n \mcdot \bar{n} \big)
\nn\\
&
\times
\left| \parbox[h]{0.28\textwidth}{
\includegraphics[width=0.25\textwidth]{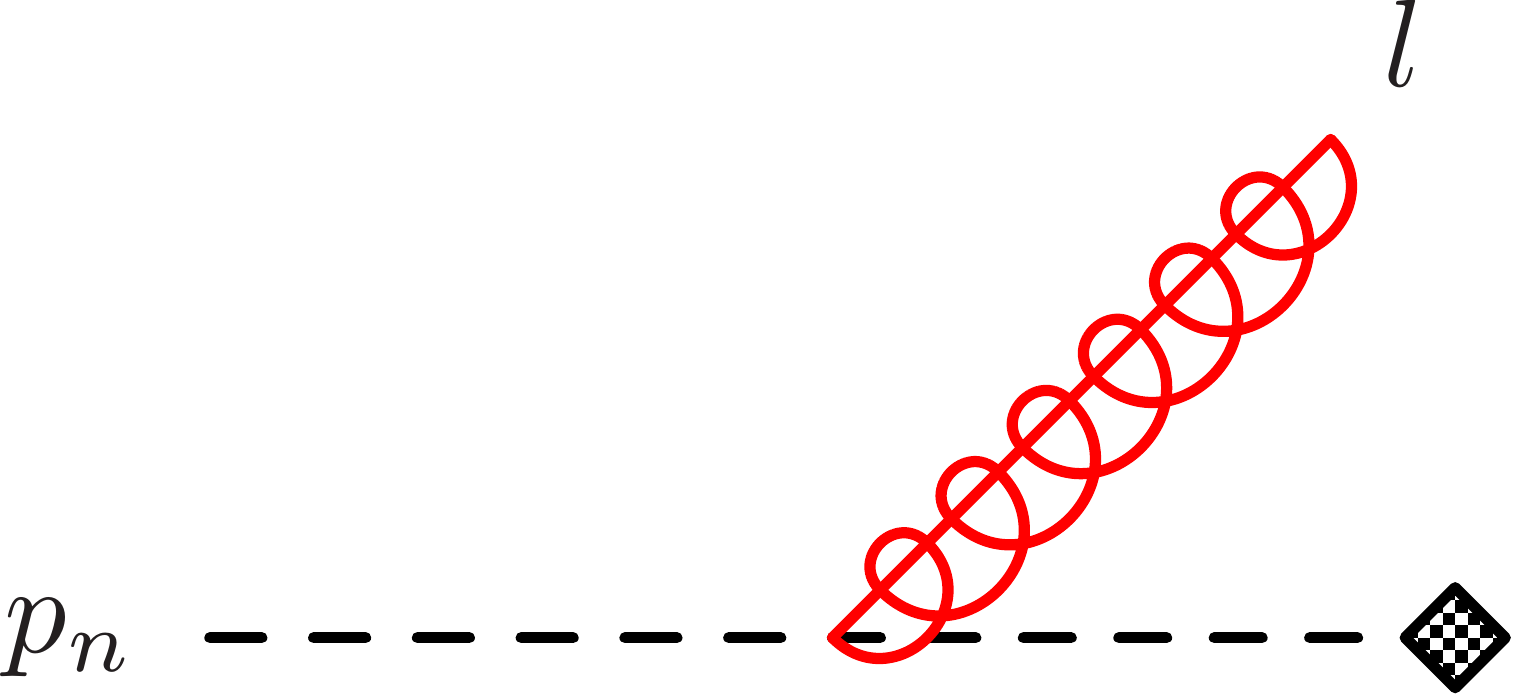} 
}
+ 
\parbox[h]{0.28\textwidth}{
\includegraphics[width=0.25\textwidth]{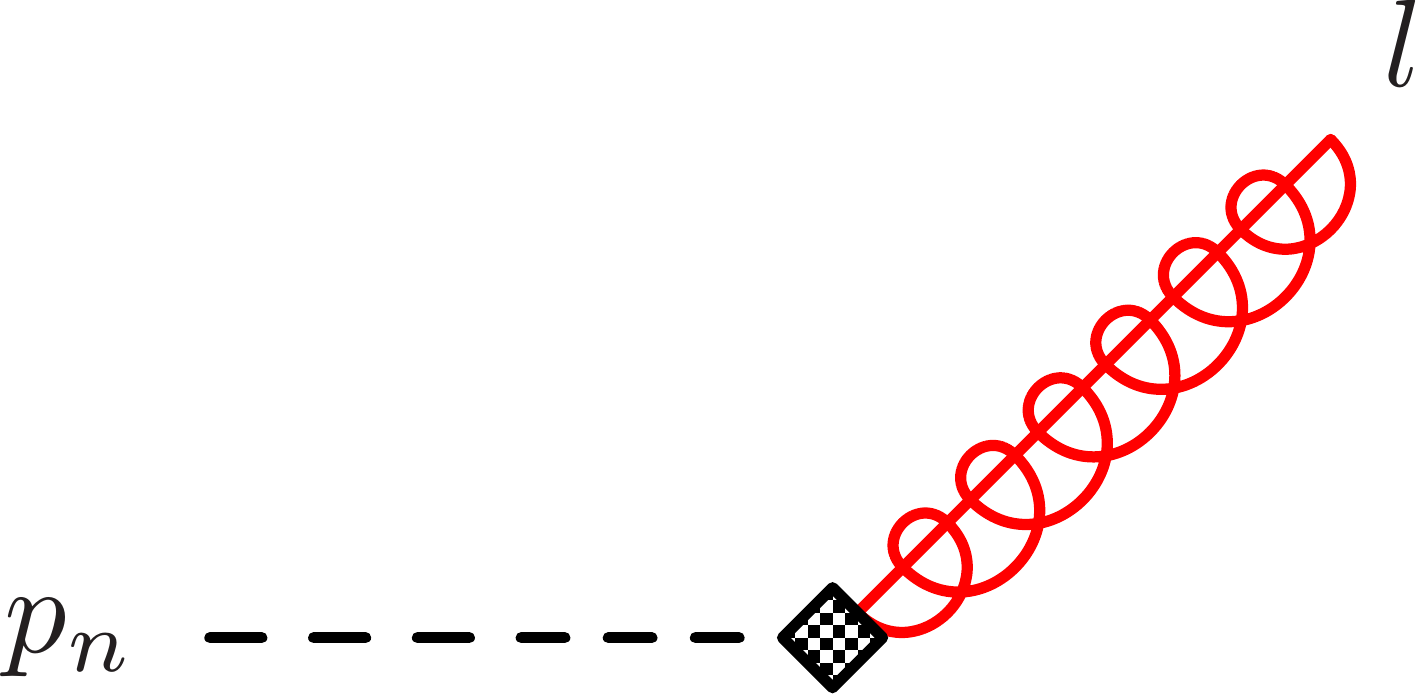} }
\right|^2 \,.
\end{align}
We also need the gluon-to-quark coefficient function at this order. 
The quark beam function has been calculated through to two loops~\cite{1401.5478}.
We quote the result up to one-loop here
\begin{align}
  \label{eq:29}
  \cI_{qq}(t, z, \mu) = &\delta(t) \delta(1-z) + \frac{\as}{2 \pi} C_F
  \left\{ 2\left[\frac{\ln (t/\mu^2)}{t} \right]_\star^{[t,\mu^2]} \delta(1-z) +
  \left[\frac{1}{t} \right]_\star^{[t,\mu^2]} \frac{(1+z^2)}{[1-z]_+}
\right.
\nn\\
&\left. 
+ \delta(t) \left[\frac{(1+z^2)}{[1-z]_+} - \frac{\pi^2}{6} \delta(1-z)
  + \left(1-z- \frac{1+z^2}{1-z} \ln z \right) \right] 
 \right\} \,,
\nn\\
\cI_{qg}(t,z,\mu) = & \frac{\as}{2 \pi} T_F \left\{ \left[\frac{1}{t}
  \right]_\star^{[t,\mu^2]} ( 1 - 2 z + 2 z^2) + \delta(t) \left[
                     (1 -2 z + 2 z^2) \left( \ln\frac{1-z}{z} - 1
                      \right) + 1 \right] \right\} \,.
\end{align}
Substituting the expansion of hard, soft and beam function into the
factorization formula in Eq.~\eqref{eq:18} gives the leading power in
$\tau$ prediction for the unresolved distribution.
The dependence on $\tau$ is very simple and can be integrated out analytically.
Note that the power suppressed terms neglected in Eq.~(\ref{eq:18}) also
can be calculated analytically and used to improve convergence of the
phase-space slicing method as demonstrated in Refs.~\cite{1612.00450,1612.02911}.

For a small cut-off $\tau_{\rm cut}$, integration of the unresolved
distribution obtained from the factorization formula results in large
logarithmic dependence on the cut-off.  
For sufficiently small cut-off, the large cut-off 
dependence is to be canceled by the resolved contribution, up to Monte-Carlo
integration uncertainty.
The resolved contribution, as its name
suggests, is free of infrared singularities at NLO. 
At NNLO, the
resolved contribution contains sub-divergences. 
These sub-divergences
cannot be resolved by our resolution variable $\tau$.
They must be canceled using other methods.
Fortunately, the infrared structure of
sub-divergences is lower by one order in $\as$ than the unresolved
part.  
For a NNLO calculation, we can use any existing subtraction
method to cancel the sub-divergences. 
In our calculation, we employ
the dipole subtraction formalism~\cite{hep-ph/9605323,hep-ph/0201036} to remove the
sub-divergences.  
We also need the one-loop amplitudes for charm quark production
with an additional parton, and tree-level amplitudes for charm quark production
with two partons.  
We extract the former from Ref.~\cite{Campbell:2005bb}; for the later we use
\texttt{HELAS}~\cite{Murayama:1992gi}.
The calculations in resolved region have been cross checked with Gosam~\cite{1404.7096} 
and Sherpa~\cite{0811.4622} and 
full agreement are found.

\subsection{Numerical results}
We move to numerical results for the reduced cross sections of charm-quark production in DIS of neutrino on iron. 
We use CT14 NNLO PDFs~\cite{Dulat:2015mca} with $n_f=3$ active quark
flavors and the associated strong coupling constant by default.
We use a pole mass $m_c=1.4$ GeV for the charm quark,
and CKM matrix elements $|V_{cs}|=0.975$ and $|V_{cd}|=0.222$~\cite{Beringer:1900zz}.
The renormalization and factorization scales are set to $\mu_0=\sqrt{Q^2+m_c^2}$ unless
otherwise specified.
We choose a phase-space slicing parameter of $\tau_{\rm cut}=10^{-3}$ which
is found to be small enough to neglect the power corrections~\cite{1601.05430}.

%FIGURE
\begin{figure}[!h]
  \begin{center}
  \includegraphics[width=0.4\textwidth]{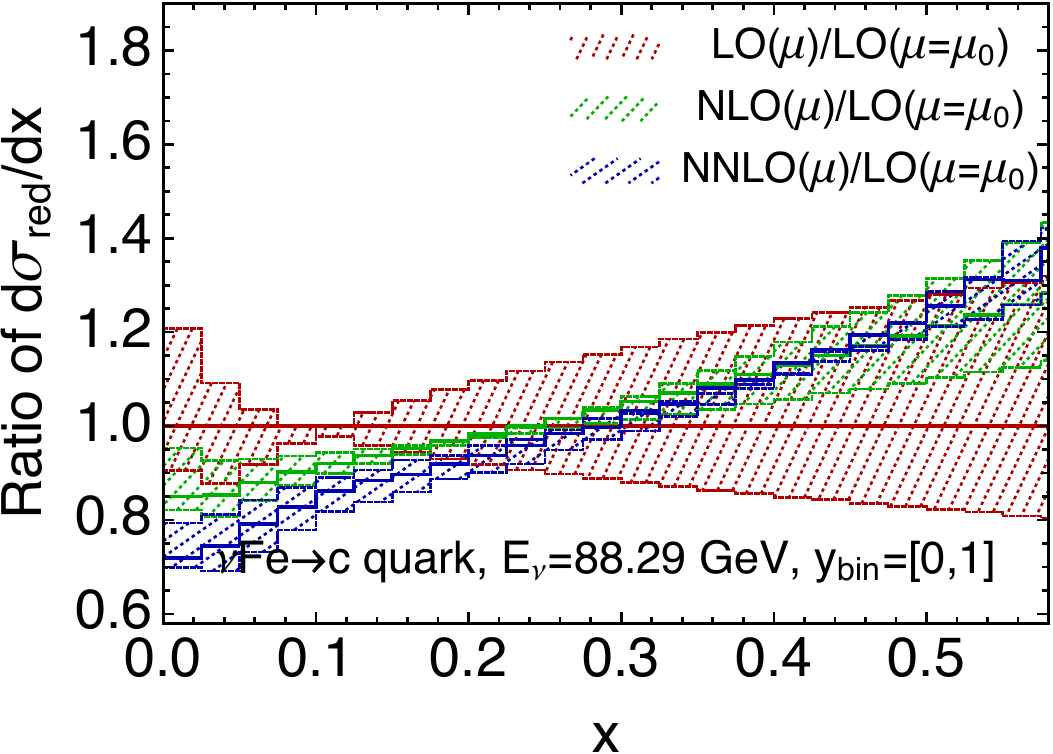}
  \hspace{0.3in}
  \includegraphics[width=0.4\textwidth]{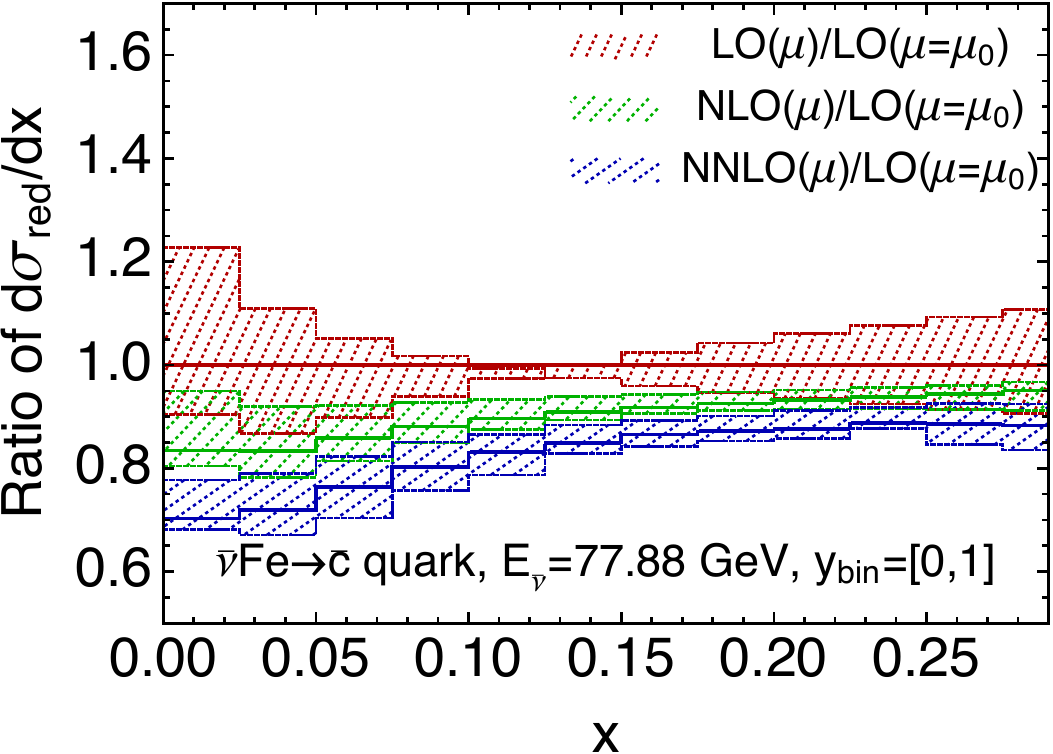}
  \end{center}
  \vspace{-2ex}
  \caption{\label{fig:scale}
  QCD predictions including scale variations at different orders
  for a differential reduced cross section
  in Bjorken $x$ for charm (anti-)quark production from (anti-)neutrino scattering
  on iron target.  
  }
\end{figure}

In Fig.~\ref{fig:scale} we show the QCD corrections to a differential reduced
cross section in Bjorken $x$ for which the electroweak couplings have been
taken out.
We plot the NLO and NNLO predictions normalized to the LO ones for charm
(anti-)quark production from (anti-)neutrino scattering with an energy
of 88.29 (77.88) GeV on iron target.
The cross sections are integrated over the full range of inelasticity $y$.
The hatched bands represent the scale variations as calculated by varying
renormalization and factorization scale from $\mu_F=\mu_R=\mu_0/2$ to $2\mu_0$
avoiding going below the charm-quark mass.
The QCD corrections are large and negative in small and moderate $x$ regions
for charm quark production with the nominal scale choice. 
The NNLO corrections can reach about -10\% for $x$ up to 0.1 and turn to
positive for $x>0.4$.
The scale variations at LO are large in general but vanish at $x\sim 0.1$
indicating its limitation as estimate of perturbative uncertainties.          
It was found even the scale variations at NLO underestimate the
perturbative uncertainties at small and moderate $x$ regions due to
accidental cancellations as will be explained later.
The NNLO scale variations give a more reliable estimation of the
perturbative uncertainties and also show improvement at high-$x$ compared
with the NLO case.
Results are similar for charm anti-quark production which can be related
via a charge conjugate parity transformation except for the differences
of initial state PDFs.
Especially the charm quark production involves Cabibbo suppressed contributions
at tree level from $d$-valence quark which dominate at high-$x$ while only sea-quark
contributions exist for charm anti-quark production.

%FIGURE
\begin{figure}[!h]
  \begin{center}
  \includegraphics[width=0.4\textwidth]{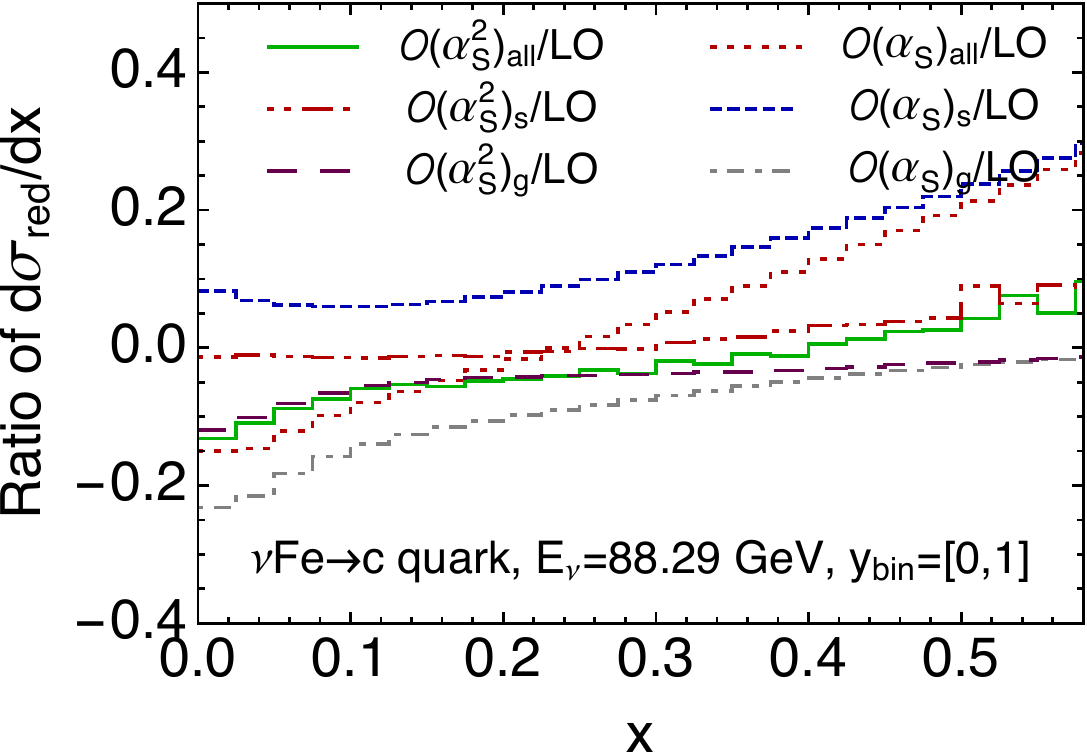}
  \hspace{0.3in}
  \includegraphics[width=0.4\textwidth]{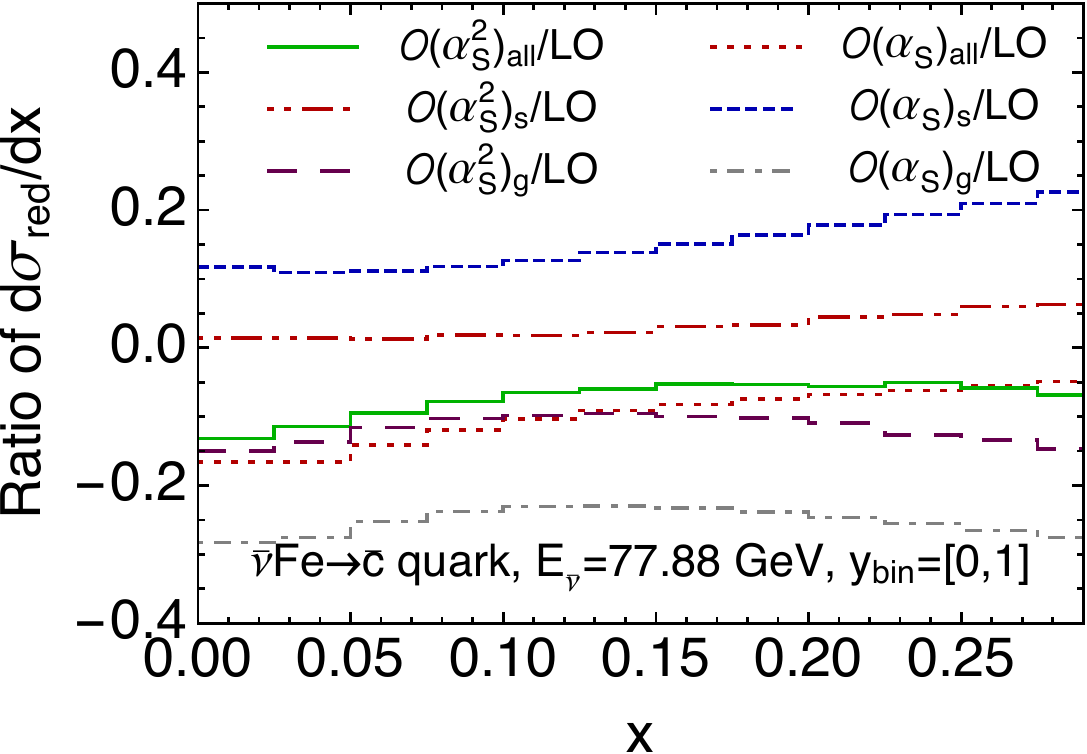}
  \end{center}
  \vspace{-2ex}
  \caption{\label{fig:subc}
  QCD corrections at different orders separated into partonic channels
  for a differential reduced cross section
  in Bjorken $x$ for charm (anti-)quark production from (anti-)neutrino scattering
  on iron target.  
  }
\end{figure}

In the small and moderate $x$ region the NNLO corrections are almost as
large as the NLO corrections.
That motivates a careful examination of the convergence of the
perturbative expansion.
In Fig.~\ref{fig:subc} we plot the QCD corrections from two main
partonic channels, i.e.,
with the strange (anti-)quark initial state, including Cabibbo
suppressed $d$($\bar d$) quark contributions, and with the gluon
initial state, for the same distribution as shown in Fig.~\ref{fig:scale}.
The right plot of Fig.~\ref{fig:subc} shows the corrections for charm
anti-quark production.
We observe a strong cancellation among the NLO corrections from
the strange anti-quark and the gluon channels starting from small-$x$
and persisting to high-$x$ region.
We regard this cancellation {\it accidental} in that it does not
arise from basic principles but is a result of several factors.
The cancellation of NLO corrections remains if instead using the NLO PDFs or
alternative NNLO PDFs e.g., MMHT2014~\cite{Harland-Lang:2014zoa} and
NNPDF3.0~\cite{Ball:2014uwa}.
A similar cancellation has also been observed in the calculation
for $t$-channel single top quark production ~\cite{1404.7116}.
The size of NNLO corrections are smaller than the NLO ones for the
individual partonic channels indicating good convergence of the
perturbative expansion.
However, the cancellation between the two channels is much mild
at NNLO which results in a net correction as large as the NLO one.
For this reason we expect the corrections from even higher orders
to be smaller than the NNLO corrections.
The left plot in Fig.~\ref{fig:subc} shows results for
charm quark production for which the situation is similar at
low-$x$ region.
At high-$x$ the correction from gluon channel flattens out
due to the smaller size of gluon PDF as comparing to $d$-valence
PDF, and the net correction is small and positive at
NNLO.

%FIGURE  
\begin{figure}[!h]
  \begin{center}
  \includegraphics[width=0.9\textwidth]{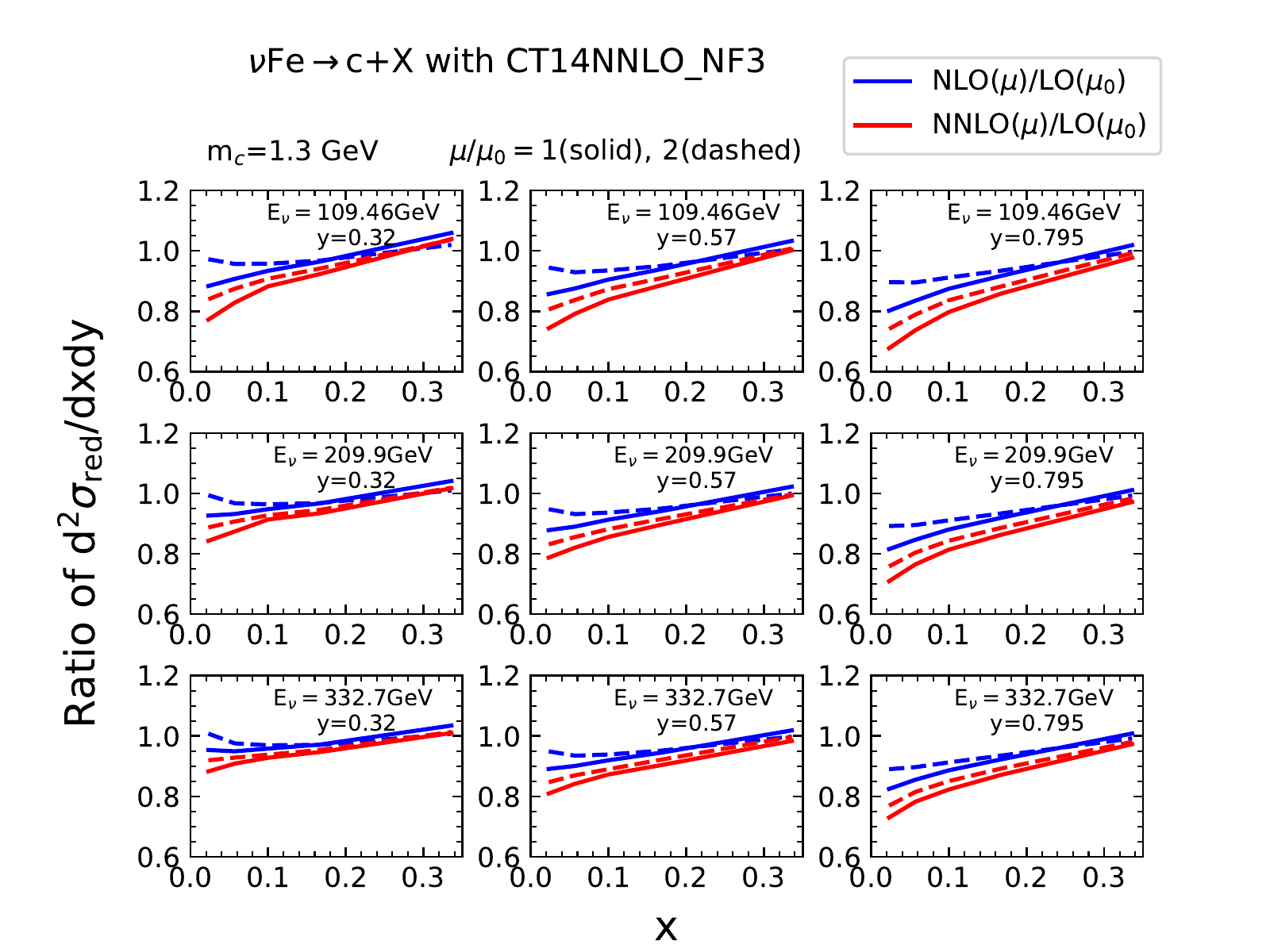}
  \end{center}
  \vspace{-2ex}
  \caption{\label{fig:kfac1}
  QCD predictions at different orders with scale choices of $\mu_0$
  and $2\mu_0$ for a double differential reduced cross section
  in Bjorken $x$ and inelasticity $y$ for charm quark production
  from neutrino scattering on iron target.  
  }
\end{figure}

%FIGURE  
\begin{figure}[!h]
  \begin{center}
  \includegraphics[width=0.9\textwidth]{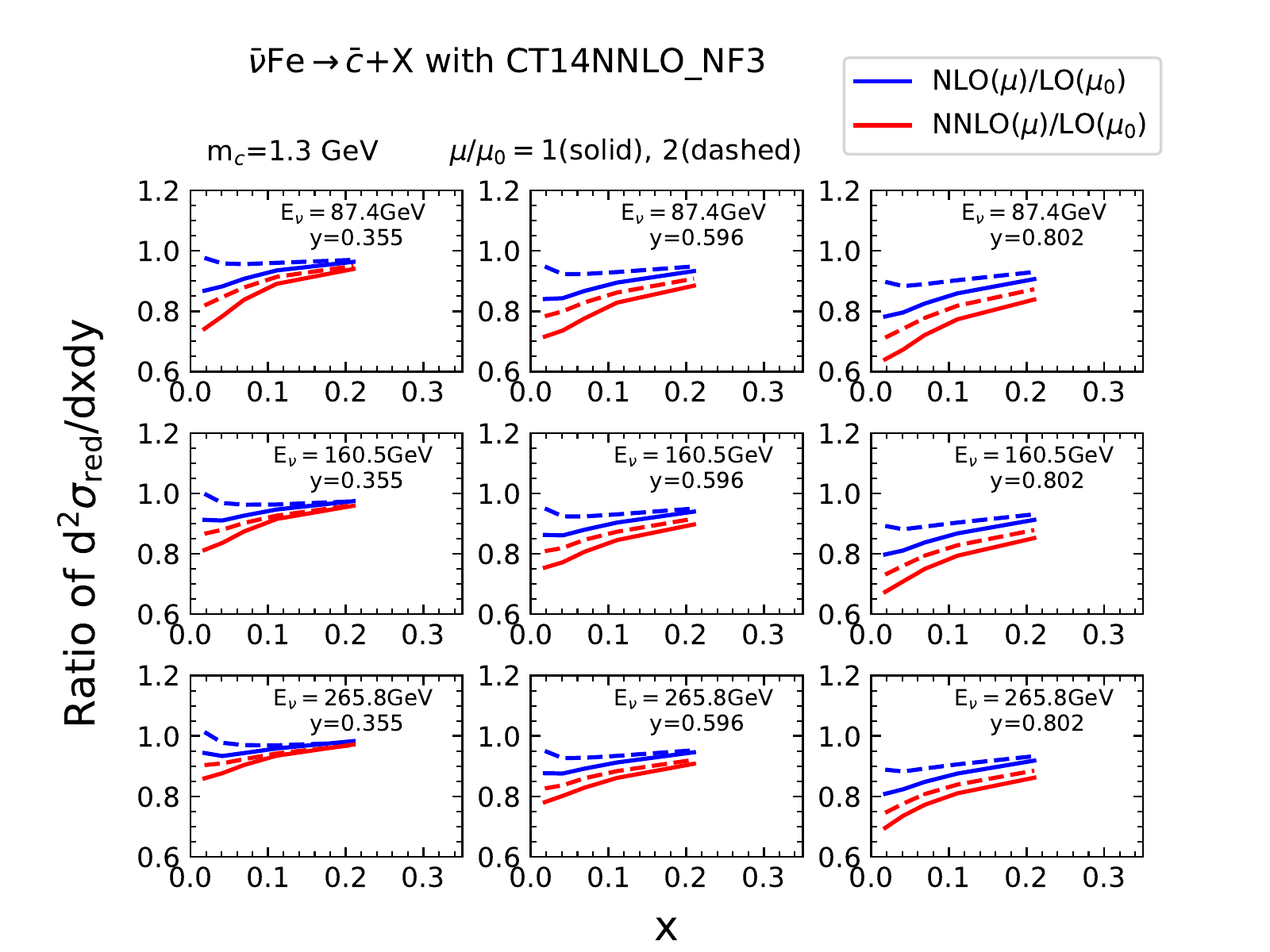}
  \end{center}
  \vspace{-2ex}
  \caption{\label{fig:kfac2}
  Similar as Fig.~\ref{fig:kfac1} for charm anti-quark production from
  anti-neutrino scattering.
  }
\end{figure}

We further calculate the double differential reduced cross sections 
in $x$ and $y$ as was measured by various experimental groups.
We choose the kinematics and neutrino energies as those in
CCFR~\cite{Goncharov:2001qe} measurement.
In Fig.~\ref{fig:kfac1} we plot ratios of various predictions to
the LO differential cross sections for charm quark production with
three different energies and each with three choices of $y$.
Here we use a charm-quark mass of 1.3 GeV.
The solid and dotted curves correspond to using scales of $\mu_0$
and $2\mu_0$.
Note the LO cross sections in the denominator are always evaluated
with the scale $\mu_0$.
For the nominal scale choice ($\mu_0$) the NNLO corrections are
about $-10$\% at $x\sim 0.02$ and a couple of percents at $x\sim 0.3$.
The size of QCD corrections increases with $y$ in low-$x$ regions.
Dependence on the beam energy is in the opposite direction and
is weaker in general.
Scale dependence of the NNLO predictions are slightly weaker than
those of the NLO predictions at small-$x$.
In moderate and large-$x$ regions the NLO predictions show a scale
dependence that is too small due to the strong cancellations mentioned
earlier.   
Fig.~\ref{fig:kfac2} shows similar results for charm anti-quark
production.
The QCD corrections are even more pronounced in this case due to
the relatively larger gluon contributions.
For $y=0.802$ the NNLO corrections can reach $-15$\% for $x\sim 0.02$
and remain $-10$\% for $x\sim 0.2$.
The same conclusion holds for the scale dependence as in the
case of charm quark production. 

%FIGURE  
\begin{figure}[!h]
  \begin{center}
  \includegraphics[width=0.9\textwidth]{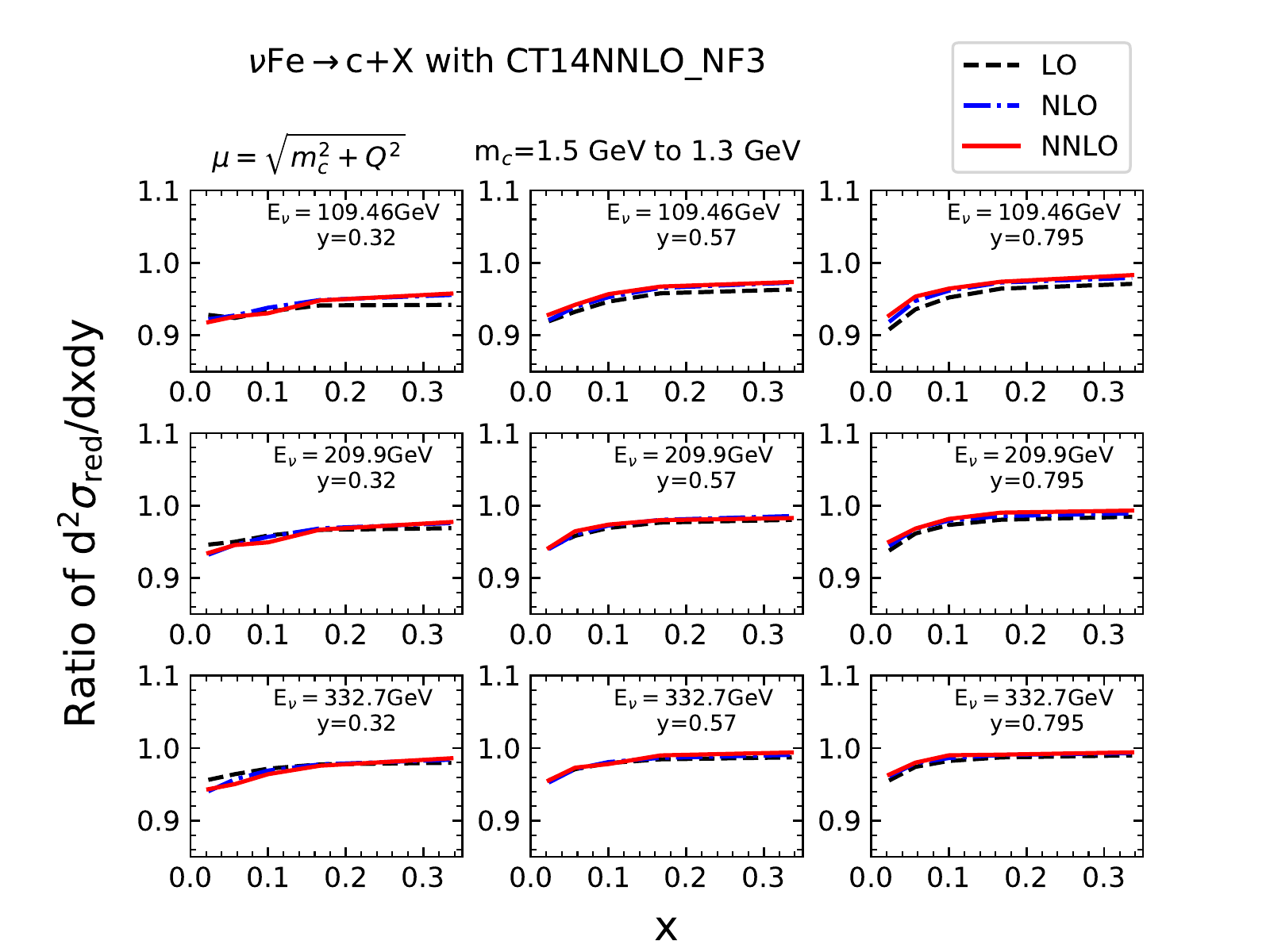}
  \end{center}
  \vspace{-2ex}
  \caption{\label{fig:mass1}
  Dependence of a double differential reduced cross section
  in Bjorken $x$ and inelasticity $y$ on the charm quark mass,
  shown in ratios of predictions with $m_c=1.5$ GeV to 1.3 GeV,  
  for charm quark production from neutrino scattering on iron target.  
  }
\end{figure}

%FIGURE  
%\begin{figure}[!h]
%  \begin{center}
%  \includegraphics[width=0.9\textwidth]{plot/CCnbCT14totmass.pdf}
%  \end{center}
%  \vspace{-2ex}
%  \caption{\label{fig:mass2}
%  %
%  Similar as Fig.~\ref{fig:mass2} for charm anti-quark production from
%  anti-neutrino scattering.
%  }
%\end{figure}

%
Since the charm quark production is usually measured at low to moderate
momentum transfers, the theoretical predictions can depend significantly
on the choice of charm-quark mass, in our case the pole mass.
Note the determination of charm-quark pole mass has an intrinsic uncertainty
of $0.1\sim 0.2$ GeV due to the renormalon ambiguity.
In Fig.~\ref{fig:mass1} we show the ratio of double
differential cross sections calculated when using a charm-quark
pole mass of 1.5 GeV to 1.3 GeV, at LO, NLO, and NNLO, for charm anti-quark production.
The results for charm quark production are similar and not shown for simplicity.
At LO the charm-quark mass dependence can be calculated easily.
The dominant part of that is known as slow rescaling~\cite{Georgi:1976vf} due
to the kinematic suppression, i.e., by replacing the momentum fraction
in evaluation of PDFs with $\xi=x(1+m_c^2/Q^2)$. 
That explains the trends we show in Fig.~\ref{fig:mass1}.
The cross sections with larger charm-quark mass are especially
suppressed in small-$x$ region and for smaller neutrino energies
where the $Q^2$ is low.
Shapes of the suppression factor with respect to $x$ are different
for different values of $y$ due to the sub-dominant dependence on
the mass from the hard matrix elements.    
The mass dependence is insensitive to higher order corrections.
The NLO predictions show a slightly weaker suppression comparing
to LO ones in general, especially at large-$x$ and smaller neutrino
energies.
Effects of NNLO corrections on the mass dependence are almost
negligible for the full kinematic range considered.  

\subsection{Heavy quark scheme}

The NNLO calculations are carried out in a fixed flavor number
scheme with $n_f=3$.
This should be the appropriate scheme for $Q\gtrsim m_c$.
For the semiinclusive charged-current (CC) DIS process we studied, at $Q \gg m_c$,
there exist logarithmic contributions of $\sim \as^n\ln^n(Q^2/m_c^2)$
due to initial-state gluon splitting into a $c\bar c$ pair in the
quasi-collinear limit.
\footnote{In this case the muon from charm decay tends to be
close to the beam, and the experimental acceptance may be different
comparing to other region of the phase space.}
In principle that needs to be resummed by using the
heavy-quark PDFs together with an appropriate general mass
variable flavor number (GM-VFN) scheme, for example, the ACOT~\cite{hep-ph/9312319},
FONLL~\cite{1001.2312}, or RT~\cite{1201.6180} schemes.
The exact ${\mathcal O}(\as^2)$ massive coefficient functions~\cite{1601.05430}
complete all ingredients needed for constructing
such a scheme like S-ACOT-$\chi$~\cite{Guzzi:2011ew} at NNLO for the charged-current
scattering.
For the kinematics where the dimuon measurements carried out,
the $Q^2$ is not too high comparing to the charm-quark mass in the
bulk of the data.
Besides, the experimental uncertainties are at least at the level of
$5\sim 10$\% for the NuTeV and CCFR measurements. 
Thus a VFN scheme is not of immediate relevance for the phenomenological study
on dimuon measurements.
We leave a formal study on the GM-VFN scheme for future publications
while  providing an estimate of the logarithmic contributions beyond
${\mathcal O}(\as^2)$ in below.

As mentioned earlier the logarithmic contributions can be
resummed effectively with a perturbative charm (anti-)quark 
PDF in $n_f=4$ scheme that follows a DGLAP evolution,
\begin{equation}
\frac{df^{(n_f=4)}_c(x, \mu^2)}{d\ln\mu^2}=\sum_{i=q,\bar q,c, \bar c, g}P_{ci}(x,\as(\mu^2))
\otimes f^{(n_f=4)}_i(x,\mu^2),
\end{equation}
where $P_{ij}$ is the DGLAP splitting function with dependence on $n_f$ suppressed,
$\mu$ is the factorization scale.
The exact results for $P_{ij}$ are known up to three loops~\cite{hep-ph/0403192,hep-ph/0404111}.  
The charm-quark PDF at arbitrary scales can be derived from the boundary
conditions at $\mu=m_c$ by evolving upward.
Note that starting at ${\mathcal O}(\as^2)$ the charm-quark PDF has
a small discontinuity at $\mu=m_c$.
We can expand the charm-quark PDF in the strong coupling constant,
\begin{align}\label{eq:expa}
f_c^{(n_f=4)}(x, \mu^2)&=\Delta^{(2)}+\left({\as(m_c^2)\over 2\pi}\right)
 \left\{L(P^{(0)}_{cg}\otimes f^{(n_f=4)}_g(x,m_c^2))\right\} 
  + \left({\as(m_c^2)\over 2\pi}\right)^2\nonumber \\ 
  &\Big\{
	 L(\sum_i P^{(1)}_{ci}\otimes f^{(n_f=4)}_i(x,m_c^2))
	 +{L^2\over 2}(\sum_i P^{(0)}_{cg}\otimes P^{(0)}_{gi}\otimes f^{(n_f=4)}_i(x,m_c^2) \nonumber \\
	 & -\beta_0 P^{(0)}_{cg}\otimes f^{(n_f=4)}_g(x,m_c^2))\Big\}+{\mathcal O}(\as^3),
\end{align} 
where $\Delta^{(2)}$ is of ${\mathcal O}(\as^2)$ due to the discontinuity
crossing the heavy-quark threshold and $L=\ln(\mu^2/m^2_c)$. 
It is understood that the strong coupling constant, the one- and two-loop
splitting functions $P^{(0,1)}_{ij}$, and the one-loop $\beta$ function in
Eq.~(\ref{eq:expa}) are all evaluated with $n_f=4$.
We can translate the strong coupling constant and PDFs with
$n_f=4$ to those with $n_f=3$ via matching at the charm-quark threshold~\cite{hep-ph/9612398}.
Furthermore, we can expand them in $\as(\mu^2)$ instead.
We arrive at an expanded solution, 
\begin{align}\label{eq:expb}
f_c^{(n_f=4)}(x, \mu^2)&=\Delta^{(2)}+\left({\as(\mu^2)\over 2\pi}\right)
 \left\{L(P^{(0)}_{cg}\otimes f^{(n_f=3)}_g(x,\mu^2))\right\} 
  + \left({\as(\mu^2)\over 2\pi}\right)^2\nonumber \\ 
  &\Big\{
	 L(\sum_i P^{(1)}_{ci}\otimes f^{(n_f=3)}_i(x,\mu^2))
	 -{L^2\over 2}(\sum_i P^{(0)}_{cg}\otimes P^{(0)}_{gi}\otimes f^{(n_f=3)}_i(x,\mu^2) \nonumber \\
	 & -\beta_0 P^{(0)}_{cg}\otimes f^{(n_f=3)}_g(x,\mu^2))\Big\}+{\mathcal O}(\as^3),
\end{align} 
with $\beta$ functions and splitting functions for $n_f=4$.
Those ${\mathcal O}(\as)$ and ${\mathcal O}(\as^2)$ logarithmic contributions
have already been captured by our NLO and NNLO calculations respectively.
Differences of the evolved charm-quark PDF and the expansion in
Eq.~(\ref{eq:expb}) can serve as an estimate of the remaining logarithmic
contributions at higher orders. 

%FIGURE  
\begin{figure}[!h]
  \begin{center}
  \includegraphics[width=0.8\textwidth]{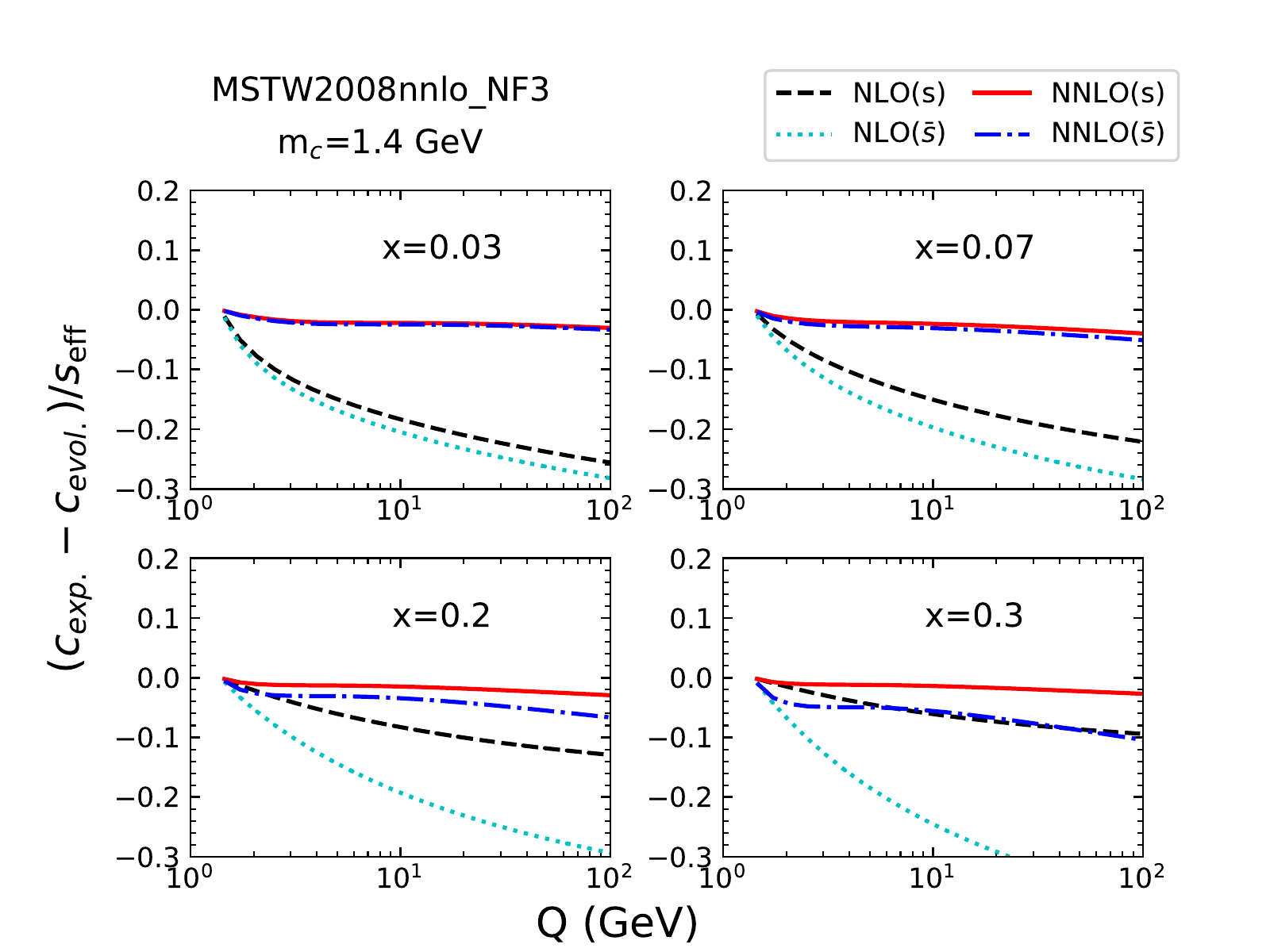}
  \end{center}
  \vspace{-2ex}
  \caption{\label{fig:vfn}
  Differences of an evolved charm-quark PDF $c_{evol.}$
  at NNLO and the expanded solution $c_{exp.}$
  up to ${\mathcal O}(\as)$ (NLO) and ${\mathcal O}(\as^2)$ (NNLO) as a function of
  $\mu=Q$ for several $x$ values, normalized to the effective strangeness PDF $s_{eff}$.
  See text for more details.
  }
\end{figure}

In Fig.~\ref{fig:vfn} we plot the differences of an evolved charm-quark PDF $c_{evol.}$
at NNLO (with 3-loop splitting functions) and the expanded solution $c_{exp.}$
up to ${\mathcal O}(\as)$ and ${\mathcal O}(\as^2)$ as a function of
$\mu=Q$ for several $x$ values, normalized to the effective strangeness PDF $s_{eff}$
which is a combination of $s(\bar s)$ and $d(\bar d)$ PDFs.
The charm quark production cross section at LO is simply
proportional to the effective strangeness PDF with the slow rescaling.
We use the MSTW2008 NNLO PDFs~\cite{0901.0002} with $m_c=$ 1.4 GeV as an input.
For small $x$ values we can see the FFN calculation at ${\mathcal O}(\as)$
misses a large portion of the logarithmic contributions that can
reach 10-20\% of the LO charm quark production cross sections for
$Q \sim 10$ GeV.
On another hand the NNLO calculation can reproduce well the resummed contributions
with the remaining logarithmic contributions of about 2\% of the
LO cross sections for the same $Q$ values.  
Note the highest $Q$ value that the CCFR and NuTeV measurements probed
is around 10 GeV.
For large $x$ values the conclusion is similar for charm quark production.
For charm anti-quark production the charm-quark PDF has a relatively
larger weight due to the quick falling of the sea-quark PDFs.  
The contributions beyond NNLO can reach 5\% for $x=0.3$ and
$Q=10$ GeV.

\section{Fast interface}
The above calculation can not be immediately used in the global analysis of QCD due
to the time-consuming nature of NNLO calculations and the fact that the analysis
involves scan over a large number of PDF ensembles.
Indeed even the NLO calculation is inadequate for direct use in the analysis.
The PDF fitting group instead needs to rely on either K-factor approximation or
fast interface based on grid interpolations.
There have been quite some developments on fast interface for high-order perturbative
calculations, e.g., APPLgrid~\cite{Carli:2010rw}, FastNLO~\cite{Wobisch:2011ij},
aMCfast~\cite{Bertone:2014zva}, starting from NLO in QCD to NNLO most
recently~\cite{Czakon:2017dip}.
We have constructed a fast interface specialized for our calculation following
similar approaches.
First of all the PDFs at arbitrary scales can be approximated by an interpolation
on a one-dimensional grid of $x$,
\begin{equation}
f(x, \mu)=\sum_{i=0}^n f_{k+i}I_i^{(n)}\left(\frac{y(x)}{\delta y}-k\right),
\end{equation}
where we choose the interpolation variable $y(x)=x^{0.3}$ and the interpolation
order $n=4$, and $f_{j}$ is the PDF value on the $j$-th grid point.
$\delta y$ has been chosen so as to give 50 grid points between $x=1$ and a minimum
determined according to the specified kinematics.
We use a $n$-th order polynomial interpolating function $I_i^{(n)}$ and the
starting grid point $k$ is determined so as $x$ located in between the
$(k+1)$-th and $(k+2)$-th grid points.
The cross section in deep inelastic scattering can thus be expressed as
\begin{equation}\label{eq:int}
d\sigma_{bin}=\sum_{p}\sum_{m}\sum_i \left(\frac{\alpha_s(\mu)}{2\pi}\right)^m
{\mathcal B}(p,m,i)f_i,
\end{equation}
where the summation runs over different sub-channels $p$, perturbative orders $m$,
and the grid points $i$.
The interpolation coefficients ${\mathcal B}(p,m,i)$ which are independent of
the PDFs can be obtained by projecting
the event weight onto the corresponding grid points during the MC integration.  
Once those interpolation coefficients are calculated and stored,
the cross sections with any PDFs can be obtained via Eq.~(\ref{eq:int})
without repeating the time-consuming calculations of the matrix elements.

\begin{figure}[!h]
  \begin{center}
  \includegraphics[width=0.7\textwidth]{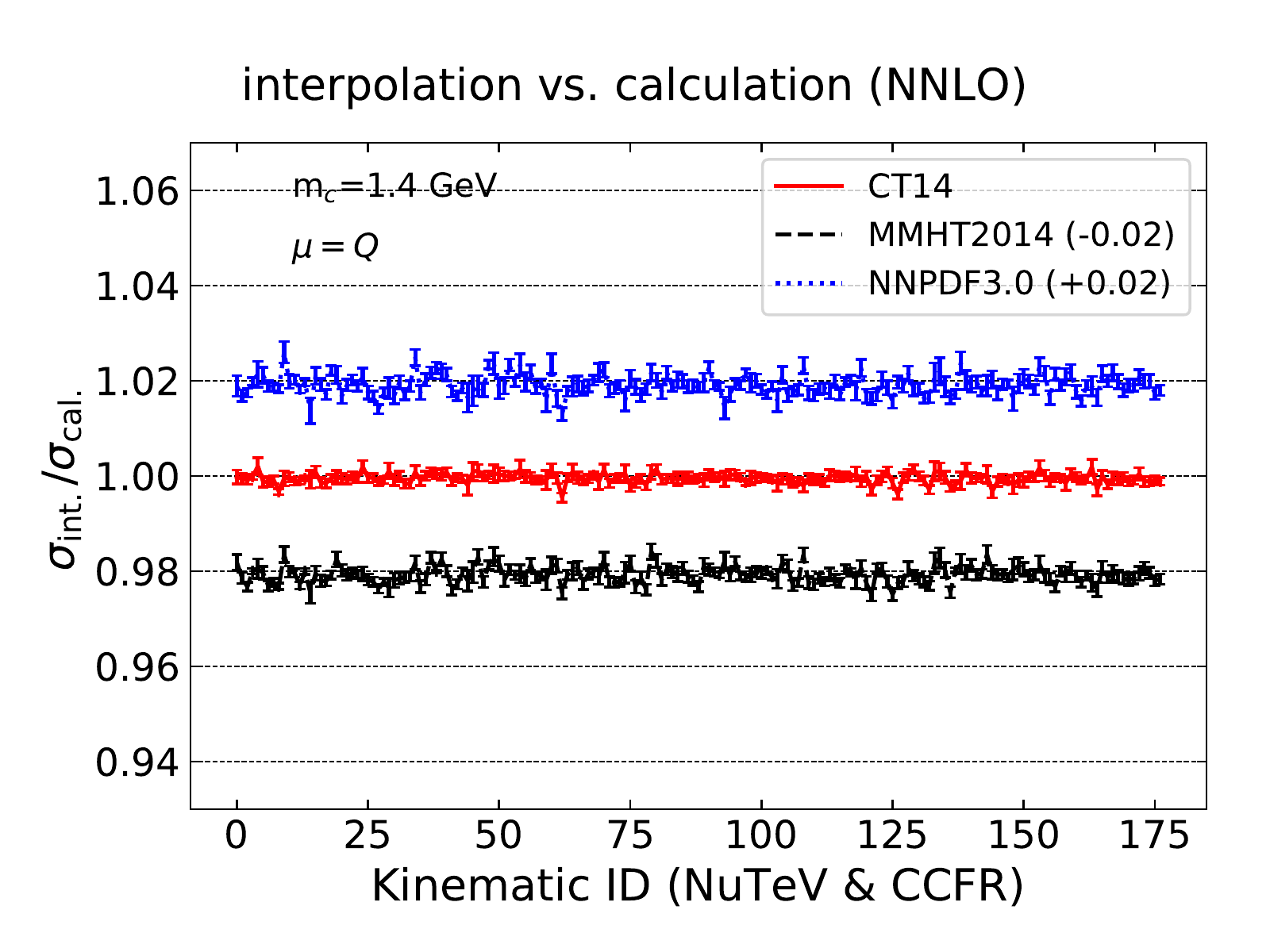}
  \end{center}
  \vspace{-2ex}
  \caption{\label{fig:int}
  Ratios of the interpolated NNLO cross sections and cross sections from direct
  calculations using CT14, MMHT2014 and NNPDF3.0 PDFs all with the grid
  generated from the same run of CT14 calculation. 
  }
\end{figure}

In Table~\ref{tab:int} we show the typical time
cost in a direct calculation and the interpolation with the NuTeV kinematics.
Also shown is the time cost for generating the interpolation grid. 
The direct calculation involves intensive MC integration
as expected, costing about 60 CPU core-hours per data point of the double differential
distribution $d^2\sigma/dxdy$ in charm-quark production with NuTeV kinematics.
The grid generation costs four times more since it requires separation
of different sub-channels.
However, with the generated grid, for any PDFs the interpolation/calculation
takes less than a millisecond.  
The precision of the interpolation are found to be around a few permille at
NNLO, smaller than the typical errors from MC integration.
In Fig.~\ref{fig:int} the solid line shows the ratio of the cross
sections from direct calculation and the fast interpolation using the grid
generated from the same run both using CT14 NNLO PDFs, for all the data
points in NuTeV and CCFR measurements with charm (anti-)quark production.  
In this case deviation of the two predictions are simply due to the interpolation
errors.
Also shown in Fig.~\ref{fig:int} are comparison of the interpolation
results for MMHT2014 and NNPDF3.0 NNLO PDFs with the grid generated from
CT14, with independent direct calculations using the same PDFs.
Here the two predictions for each PDF choice differ at most half percent due to the
MC integration errors in the direct calculations as shown by the error bars.
%
% TABLE   
\begin{table}[h!]
\centering
\begin{tabular}{c|cc}
\hline
& CPU core-hours (NLO) & CPU core-hours (NNLO) \tabularnewline
\hline
\hline
direct calculation & 0.5 & 60 \tabularnewline
\hline
grid generation & 1 & 280 \tabularnewline
\hline
interpolation & $10^{-7}$ & $10^{-7}$\tabularnewline
\hline
\end{tabular}

\caption{
  Typical time cost (in CPU core-hours) for calculation and interpolation
  of reduced cross section $d^2\sigma/dxdy$ (per data point) of charm-quark
  production with NuTeV kinematics.    
\label{tab:int}}
\end{table}

%FIGURE

\section{Impact on strange-quark distributions}
Now we move to discuss potential impact of the NNLO calculations on constraining
parton distributions, especially the strange-quark distributions, by checking
the agreements of different theories with experimental data. 
We select the NuTeV and CCFR measurements on charm (anti-)quark production
in the form of double differential cross sections $d^2\sigma/dxdy$, which
provides dominant constraints on the strange-quark distributions, e.g., in
MMHT2014 and CT14 global analyses.
The theoretical predictions used in those analyses are at NLO only.
We include the data point only with $Q^2>4\,{\rm GeV}^2$ to leave out
the region where higher-twist corrections can be potentially large.
That results in 38(33) data points for charm (anti-)quark production in NuTeV,
and 40(38) data points for charm (anti-)quark production in CCFR.
Besides, we have simply corrected the data for nuclear effects of iron
target~\cite{hep-ph/0312323,1203.1290} using a parametrization of $F_2$ ratio
measured at SLAC and NMC, instead of including more sophisticated corrections
to individual parton flavors~\cite{0709.3038,0902.4154,1112.6324,1509.00792} in
the theory calculations.
That leads to corrections on the data of 2\% at $x\sim 0.05$, -4\% at $x\sim 0.1$,
and $5\%$ at $x\sim 0.4$.
We did not include uncertainties on the nuclear corrections since the correction
itself is already small comparing to experimental errors for the $x$ range considered.
The experimental uncertainties include the total statistical and systematic
errors which are treated uncorrelated among different data point.
The total error for each data point has been scaled by square root of its effective
freedom so as a reasonable fit should have $\chi^2/N_{pt}$ of one~\cite{Mason:2006qa}.
Besides, there is an additional systematic error of 10\% assumed due to
the input of semi-leptonic decay branching ratio of charm quark when unfolding
the dimuon cross sections back to the charm production cross sections~\cite{Mason:2006qa}.
This normalization error is assumed to be fully correlated among
all data points in both NuTeV and CCFR measurements.     

In the following we first compare predictions with various PDFs to
the experimental measurement. 
Later we show how the strange quark
distributions may change if using our NNLO results instead of the
NLO ones in the PDF analyses by means of Hessian profiling~\cite{1503.05221}.    

\subsection{Theory-data agreement with various PDFs}
We considered the updated NNLO PDFs from the major PDF fitting
groups, including CT14~\cite{Dulat:2015mca}, MMHT14~\cite{Harland-Lang:2014zoa},
NNPDF3.1 (both nominal set and set with only collider data)~\cite{Ball:2017nwa}, ABMP16~\cite{Alekhin:2017kpj},
HERAPDF2.0~\cite{1506.06042}, and ATLAS-epWZ16~\cite{1612.03016}.
For the case of NNPDF, the original PDF representation of MC replicas
has been transformed into a Hessian PDF set using the MC2H package~\cite{1505.06736}. 
Note that we have used NNLO PDFs for both NLO and NNLO calculations.
In case that the PDFs with $n_f=3$ are not publicly available we evolve
the nominal PDFs by DGLAP evolution at 3-loop and with $n_f=3$ starting from
a scale below the charm quark mass threshold.

We show the fits to NuTeV and CCFR data (149 data points in total) with
NLO and NNLO predictions for various choices of PDFs in Table~\ref{tab:chi2a}.
Here through the calculations we have used a charm-quark pole mass of 1.3 GeV
and a scale of $\mu=\sqrt{Q^2+m_c^2}$ despite the fact that different
PDF groups use a charm-quark pole mass ranging from 1.3
to 1.6 GeV.\footnote{ABMP16 uses $\overline{\rm MS}$ mass as input and
treat them in the same footing as the other PDF parameters.}
In each fit we show the $\chi^2$ and the normalization shift (in unit of $1\,\sigma$ error)
of the central PDFs without and with including the full PDF uncertainties.
In the later case it gauges the overall agreement between data and theory
with both uncertainties included.     
Shift with minus sign indicates the data prefer smaller values for the cross sections.
Each pair of eigenvector PDFs corresponds to one correlated theoretical error
with symmetric Gaussian distribution when including the PDF uncertainties~\cite{1503.05221}.  
For the HERA and ATLAS fits the PDF uncertainties include those model and parametrization
uncertainties as well.
Note the $\chi^2$ shown here may not represent the actual fit quality of the
same data in their global analyses since there different predictions or input
parameters are used.	

% TABLE   
\begin{table}[h!]
\centering
\begin{tabular}{c|cc|cc}
\hline
$N_{pt}=149$ &\multicolumn{2}{c|}{NLO} & \multicolumn{2}{c}{NNLO}\tabularnewline
\hline
CT14 & 167.3(-1.0) & {\bf 130.2(1.1)} & 154.2(-0.4) &{\bf 132.9(1.3)} \tabularnewline
\hline
MMHT14 & 132.2(-1.0) & {\bf 118.6(0.1)}  & 127.7(-0.3) &{\bf 118.8(0.1)}  \tabularnewline
\hline
NNPDF3.1 & 157.8(-1.2) & {\bf 115.8(-1.0)}  & 161.3(-0.5) &{\bf 115.1(-0.6)}  \tabularnewline
\hline
ABMP16 & 189.3(-1.6) & {\bf 170.8(-0.8)}  & 170.2(-1.0) &{\bf 157.6(-0.3)}  \tabularnewline
\hline
HERAPDF2.0 & 258.4(-0.8) & {\bf 130.3(0.3)}  & 221.6(-0.1) &{\bf 132.0(0.5)}  \tabularnewline
\hline
ATLAS-epWZ16 & 352.8(-4.0) & {\bf 246.6(-2.1)}  & 321.5(-3.7) &{\bf 228.7(-1.6)}  \tabularnewline
\hline
NNPDF3.1 (collider) & 513.4(-5.1) & {\bf 118.5(-2.3)}  & 537.8(-4.8) &{\bf 114.0(-1.9)}  \tabularnewline
\hline
\end{tabular}

\caption{
  $\chi^2$ and normalization shift (in unit of $1\,\sigma$ error) of fits to
  NuTeV and CCFR charm production data with various theoretical predictions
  using $m_c=1.3$ GeV and $\mu=\sqrt{Q^2+m_c^2}$.
  The shifts are shown in brackets with minus sign indicating the data prefer
  smaller values for the cross sections.
  The numbers in bold font correspond to fits including
  the full PDF uncertainties as well. 
\label{tab:chi2a}}
\end{table}

\begin{figure}[!h]
  \begin{center}
  \includegraphics[width=0.9\textwidth]{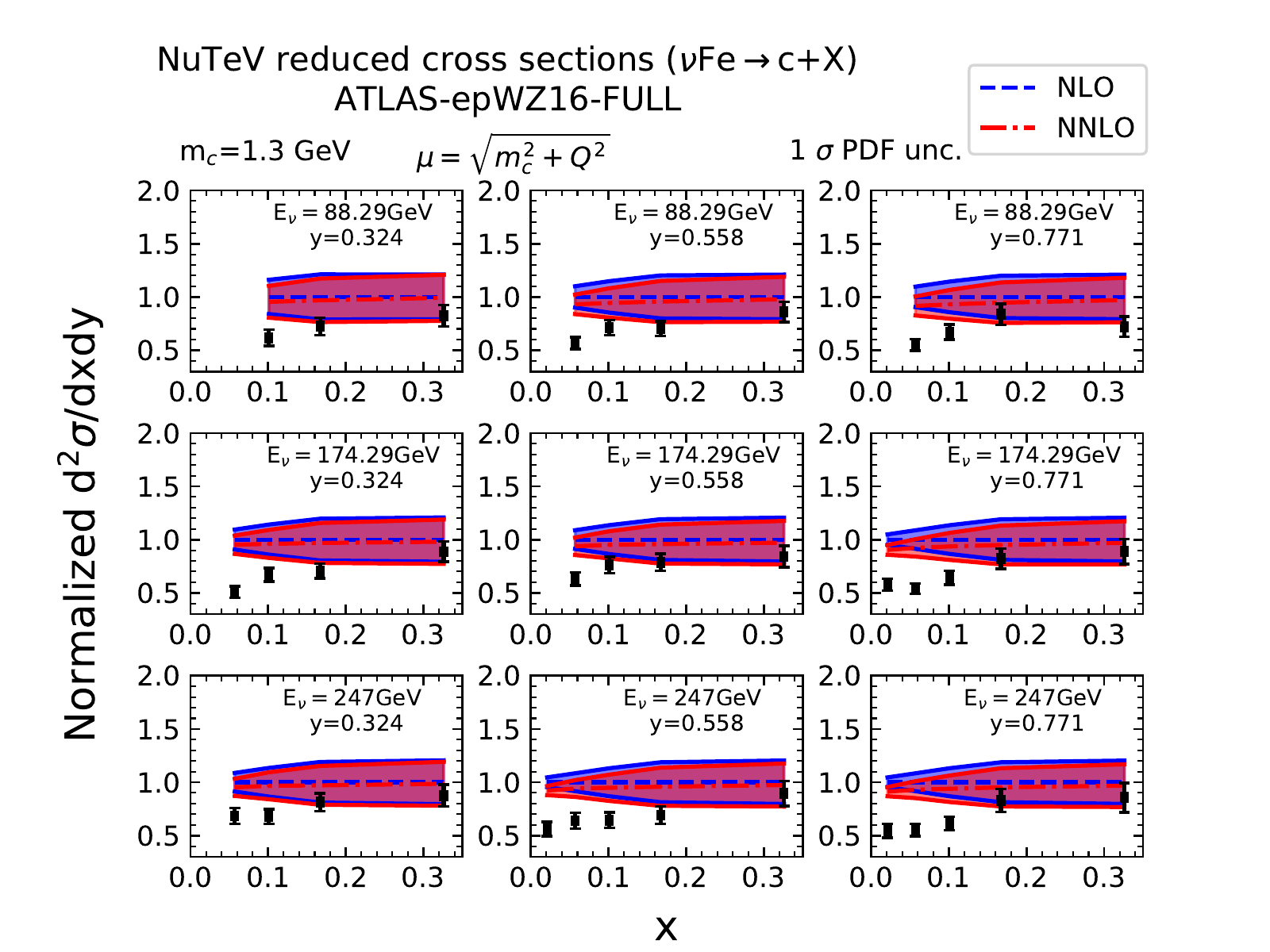}
  \end{center}
  \vspace{-2ex}
  \caption{\label{fig:dtcom_atlas16a}
  NLO and NNLO predictions on charm quark production cross sections  
  using ATLAS-epWZ16 NNLO PDFs comparing with the NuTeV measurement.
  Also shown are the 1 $\sigma$ PDF uncertainties.
  The experimental data have been corrected for nuclear effects.
  The error bars represent the total experimental uncertainties
  rescaled by square root of the effective degree of freedoms.
  A 10\% normalization uncertainty due to the semi-leptonic
  decay BR of charm quark is not shown here. 
  }
\end{figure}

\begin{figure}[!h]
  \begin{center}
  \includegraphics[width=0.9\textwidth]{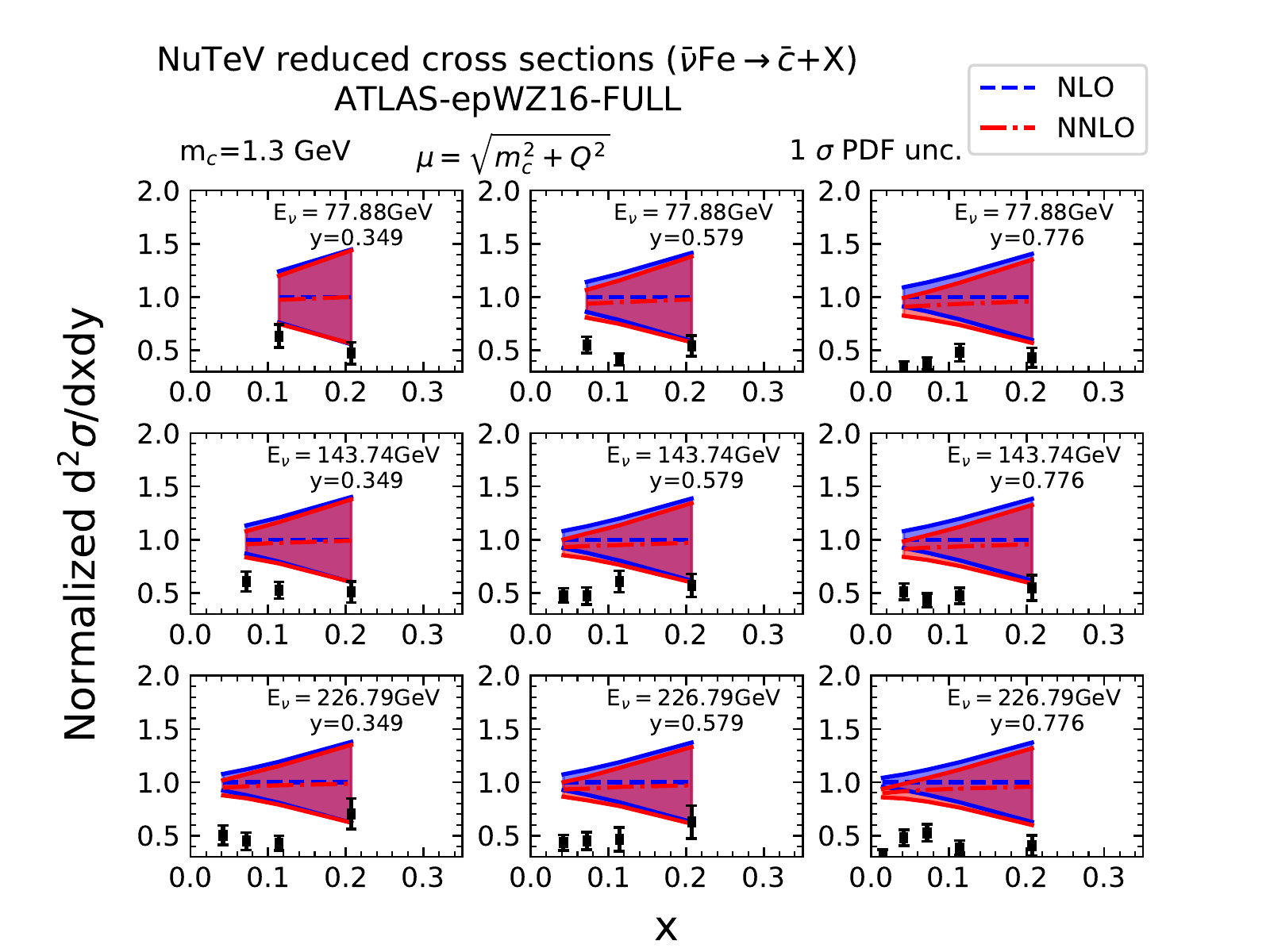}
  \end{center}
  \vspace{-2ex}
  \caption{\label{fig:dtcom_atlas16b}
  Similar as Fig.~\ref{fig:dtcom_atlas16a} for charm anti-quark production
  using ATLAS-epWZ16 NNLO PDFs.
  }
\end{figure}

The PDFs shown in Table~\ref{tab:chi2a} fall into two distinct groups, those without
including any dimuon data in the PDF analysis, namely the HERA, ATLAS, and NNPDF
collider-only PDFs, and the others with dimuon data, either from NuTeV, CCFR,
CHORUS, or NOMAD. 
The $\chi^2/N_{pt}$ are about one for PDFs in the second group as all the dimuon data
consistently prefer a suppressed strangeness as discussed in the introduction
section.
Interestingly, CT14, MMHT14 and NNPDF3.1 show very similar results on $\chi^2$ and
also the normalization shift. 
The NNLO predictions without PDF uncertainties give slightly smaller $\chi^2$ in
general for the specified mass
and scale as comparing to NLO,
though those PDFs are fitted with NLO or approximate NNLO predictions.
For PDFs from collider-only data, the fits are rather poor if not taking into
account the PDF uncertainties.
One reason is the ATLAS W/Z data do prefer larger central values
of the strange-quark PDFs.
With the PDF uncertainties included, the $\chi^2/N_{pt}$ have been reduced to
below one for HERA and NNPDF collider-only PDFs indicating consistency of their
PDFs and the dimuon data once both uncertainties are considered.
The situation is different for the ATLAS PDFs where $\chi^2/N_{pt}$ is still
about 1.5 even for the NNLO predictions.
That can be further visualized by a direct comparison of theory and data
as in Figs.~\ref{fig:dtcom_atlas16a} and~\ref{fig:dtcom_atlas16b} for
NuTeV charm quark and anti-quark production respectively.
Most of the data points lie far outside the PDF error bands with a non-trivial
shape dependence.
The PDF uncertainties of charm quark cross sections are smaller than
the ones of charm anti-quark in general since the former also involves contribution
from $d$ valence quark which is better constrained than sea quarks.
We conclude the ATLAS PDFs can not describe the dimuon data well and
the NNLO calculations can only bring in limited improvement. 

%FIGURE  

%FIGURE

We further compare predictions from the same PDF sets but with
different scale and charm quark mass inputs as shown in Tables~\ref{tab:chi2b}
and~\ref{tab:chi2c}. 
From Table~\ref{tab:chi2b} we can see that by using a scale of
twice of the nominal choice the agreement between NLO predictions
and data deteriorates, especially in the case of without PDF
uncertainties.
In comparison the $\chi^2$ for NNLO predictions are less sensitive to the
change of scale for both with and without PDF uncertainties.
The cross sections are reduced especially at small-$x$ when using a
larger charm-quark mass of 1.4 GeV as shown already in Fig.~\ref{fig:mass1}.
As show in Table~\ref{tab:chi2c} that leads to a smaller $\chi^2$ at NLO in
general comparing with Table~\ref{tab:chi2a} when no PDF uncertainties are included.
At NNLO the $\chi^2$ can either decrease or increase depending on the
PDFs considered indicating a different preference of charm-quark mass
at NNLO comparing with NLO for certain PDFs.
The $\chi^2$ for the ATLAS PDFs are reduces as well.
However, in both cases
the $\chi^2$ are still over 200 for predictions with the ATLAS PDFs.

% TABLE   
\begin{table}[h!]
\centering
\begin{tabular}{c|cc|cc}
\hline
$N_{pt}=149$ &\multicolumn{2}{c|}{NLO} & \multicolumn{2}{c}{NNLO}\tabularnewline
\hline
CT14 & 196.1(-1.3) & {\bf 131.6(1.2)} & 160.3(-0.7) &{\bf 130.5(1.3)} \tabularnewline
\hline
MMHT14 & 152.7(-1.3) & {\bf 123.1(0.0)}  & 127.1(-0.6) &{\bf 117.7(0.2)}  \tabularnewline
\hline
NNPDF3.1 & 163.1(-1.5) & {\bf 119.2(-1.2)}  & 153.2(-0.8) &{\bf 114.4(-0.7)}  \tabularnewline
\hline
ABMP16 & 223.5(-1.8) & {\bf 197.1(-1.1)}  & 180.6(-1.3) &{\bf 161.8(-0.6)}  \tabularnewline
\hline
HERAPDF2.0 & 308.4(-1.2) & {\bf 130.3(0.5)}  & 238.9(-0.5) &{\bf 130.2(0.5)}  \tabularnewline
\hline
ATLAS-epWZ16 & 391.5(-4.1) & {\bf 271.4(-2.4)}  & 339.7(-3.8) &{\bf 239.4(-1.8)}  \tabularnewline
\hline
NNPDF3.1 (collider) & 487.7(-5.1) & {\bf 124.1(-2.6)}  & 521.0(-5.0) &{\bf 116.4(-2.0)}  \tabularnewline
\hline
\end{tabular}

\caption{
  Similar as Table~\ref{tab:chi2a} but with $\mu=2\sqrt{Q^2+m_c^2}$.
\label{tab:chi2b}}
\end{table}

% TABLE   
\begin{table}[h!]
\centering
\begin{tabular}{c|cc|cc}
\hline
$N_{pt}=149$ &\multicolumn{2}{c|}{NLO} & \multicolumn{2}{c}{NNLO}\tabularnewline
\hline
CT14 & 158.2(-0.8) & {\bf 131.1(1.0)} & 150.5(-0.1) &{\bf 134.1(1.3)} \tabularnewline
\hline
MMHT14 & 128.2(-0.8) & {\bf 118.9(0.0)}  & 129.4(-0.1) &{\bf 119.6(0.1)}  \tabularnewline
\hline
NNPDF3.1 & 156.6(-1.0) & {\bf 115.9(-0.9)}  & 166.4(-0.3) &{\bf 115.5(-0.5)}  \tabularnewline
\hline
ABMP16 & 177.1(-1.4) & {\bf 162.6(-0.7)}  & 163.2(-0.8) &{\bf 153.2(-0.1)}  \tabularnewline
\hline
HERAPDF2.0 & 240.9(-0.6) & {\bf 130.5(0.2)}  & 209.2(0.2) &{\bf 132.6(0.5)}  \tabularnewline
\hline
ATLAS-epWZ16 & 332.8(-3.9) & {\bf 234.4(-2.0)}  & 303.5(-3.5) &{\bf 218.9(-1.5)}  \tabularnewline
\hline
NNPDF3.1 (collider) & 527.0(-5.0) & {\bf 116.4(-2.2)}  & 553.7(-4.8) &{\bf 110.2(-1.9)}  \tabularnewline
\hline
\end{tabular}

\caption{
  Similar as Table~\ref{tab:chi2a} but with $m_c=1.4$ GeV and $\mu=Q$.
\label{tab:chi2c}}
\end{table}

\subsection{Hessian profiling of PDFs}
One main motivation of this paper is to investigate impact of the NNLO
calculations on extraction of the strange quark PDFs in global analyses including
the dimuon data.
Precisely we would like to see how the outcome strange quark PDFs
change when using NNLO predictions instead of the NLO ones,
as only NLO predictions are available in previous PDF fits.    
That could be done by individual PDF groups using the fast interface
and grids presented in this paper.
Alternatively we can estimate the possible shift of the PDFs
by means of Hessian profiling~\cite{1503.05221}.
In Hessian profiling the PDF parameters are assumed to follow a
prior multi-Gaussian distribution.
That corresponds effectively to a parabolic shape of the prior $\chi^2$
around the central PDF and with $\Delta\chi^2=1$ when reaching the
$1\,\sigma$ error in each eigenvector direction.
The $\chi^2$ of any new data set to be included also forms a
parabola in the PDF parameter space under linear approximation.
The profiled PDFs are thus determined according to profiles of
the total $\chi^2$, e.g., by minimization of the
total $\chi^2$ for the central value and $\Delta\chi^2=1$ for
the PDF uncertainties.
We stress here the Hessian profiling can only serve as an estimation
of the effects of new data or theory on the PDFs since an actual
PDF fit involves further complexities due to e.g., parametrization
dependence and the requirement of tolerance conditions. 

%FIGURE
\begin{figure}[!h]
  \begin{center}
  \includegraphics[width=0.6\textwidth]{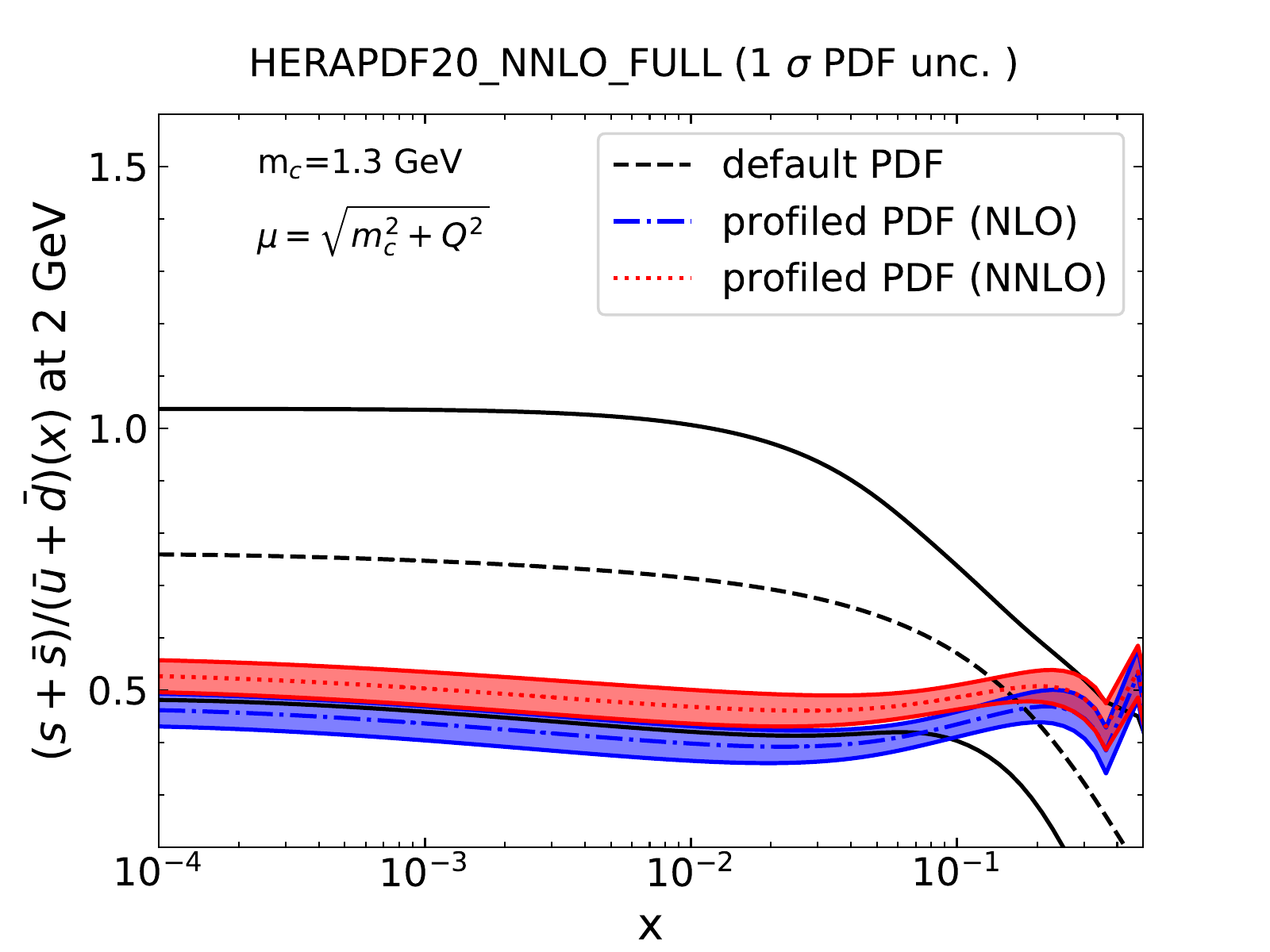}
  \end{center}
  \vspace{-2ex}
  \caption{\label{fig:prof_hera}
  PDF ratio of strange to non-strange sea-quarks at 2 GeV as a function of $x$
  for the HERAPDF2.0 NNLO PDFs.
  The solid black lines indicate the upper/lower uncertainties of original
  PDFs.
  The colored bands represent the central value and uncertainty
  after profiling using the NuTeV and CCFR charm (anti-)quark production
  data with the NLO and NNLO predictions. 
  }
\end{figure}

We start with the HERAPDF2.0 NNLO PDFs which does not include any dimuon
data but rather implements certain model constraints on the strangeness fraction 
and shape.
We show the PDF ratio of strange to non-strange sea-quarks, $R_s$ at a scale of 2 GeV
in Fig.~\ref{fig:prof_hera}.
We can see a moderate suppression of the strangeness in small to intermediate
$x$ region comparing to the $u$ and $d$ sea-quarks and a rapid falloff at
large $x$.
The PDF uncertainties as indicated by the solid black lines are large,
more than 30\% in the entire $x$ range,
which include the experimental, model, and parametrization uncertainties.
The colored bands in Fig.~\ref{fig:prof_hera} are for the profiled PDFs
with the NuTeV and CCFR charm (anti-)quark production data together using the NLO
and NNLO theoretical predictions.
It is clear that the dimuon data prefers an even suppressed strangeness with
$R_s$ of about 0.5 in the full range of $x$.
The profiled PDFs lie at the lower edge of the $1\, \sigma$ error of the
original PDFs indicating reasonable agreement between original PDFs
and the dimuon data as already seen in Table~\ref{tab:chi2a}.
The profiled PDFs have a much smaller uncertainties on $R_s$ than the original
PDFs as one expect.
We notice that the PDF uncertainties are also reduced significantly in the small $x$
region $10^{-4}-10^{-2}$ which are beyond the coverage of the dimuon
data.
That is possibly due to the restricted parametrization form of strange
quark PDFs used in the HERA PDF analysis.
Importantly we found the NNLO predictions prefer higher values
of $R_s$ than the NLO ones, in this case well above the $1\,\sigma$
error band of the later.
That can be understood since the NNLO corrections are negative in the bulk
of the data.

%FIGURE
\begin{figure}[!h]
  \begin{center}
  \includegraphics[width=0.6\textwidth]{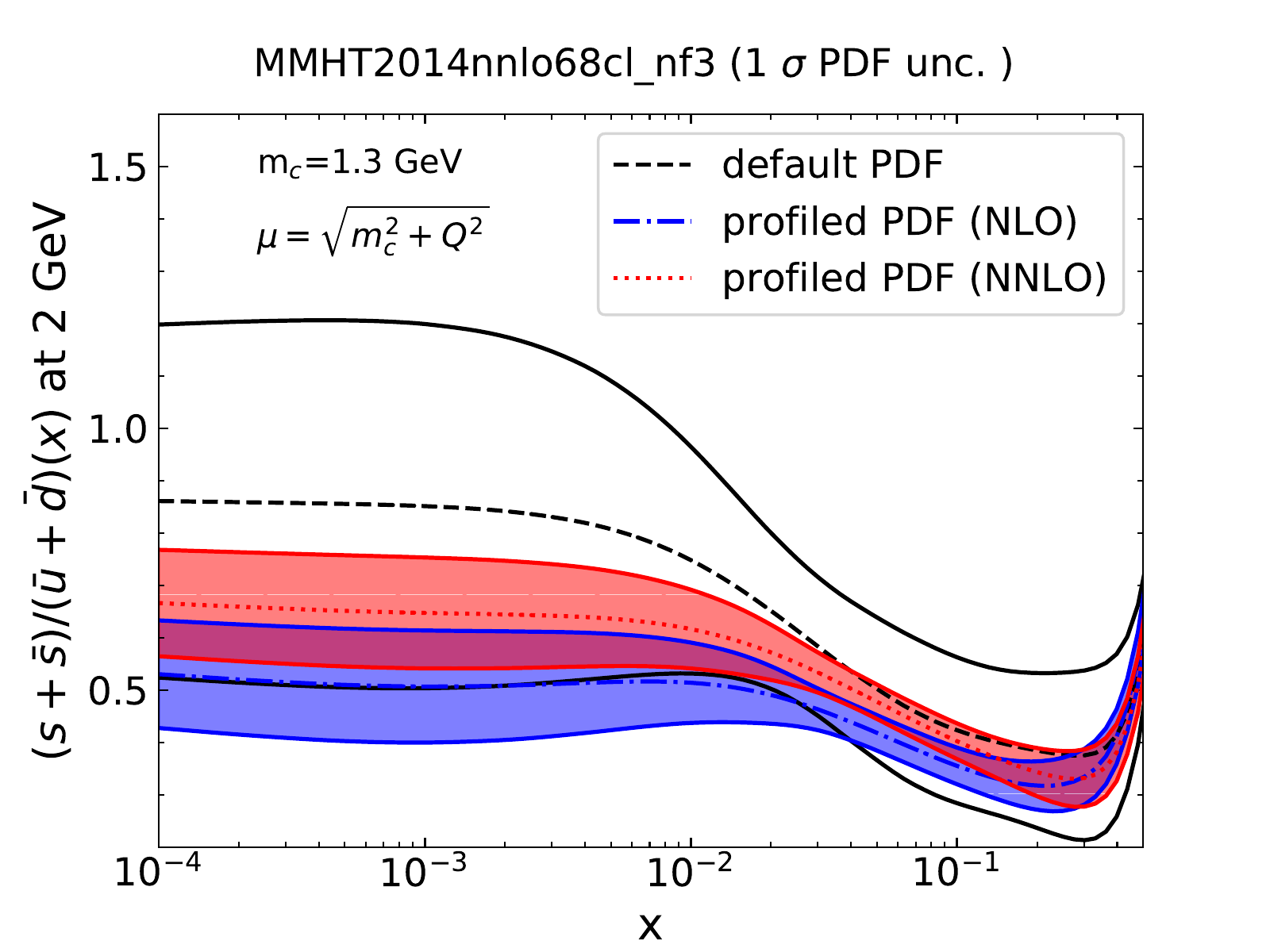}
  \end{center}
  \vspace{-2ex}
  \caption{\label{fig:prof_mmht14}
  Similar as Fig.~\ref{fig:prof_hera} for profiling of 
  the MMHT2014 NNLO PDFs.
  }
\end{figure}

We perform another profiling study with the MMHT2014 NNLO PDFs as shown
in Fig.~\ref{fig:prof_mmht14}.
Noted that since the MMHT2014 analysis already includes above dimuon data,
the study here only means for checking impact of the NNLO corrections.
We can see the NNLO predictions prefer larger strangeness than NLO predictions
for $x$ up to a few times 0.1 and by a similar amount as in
Fig.~\ref{fig:prof_hera}.
The shift of central values of the NLO profiled PDFs comparing to the original
PDFs, though still within the PDF error band, is due to several facts.
In the MMHT2014 fits~\cite{Harland-Lang:2014zoa} they use a charm-quark
pole mass of 1.4 GeV and a semi-leptonic
decay branching ratio of charm quark that is 7\% lower than the one extracted
by NuTeV and CCFR, both of which lead to an increase of the strange-quark PDFs.
Besides, there are also LHC data in the MMHT analysis that pull the ratio further up.
The uncertainties are largely reduced in the profiled PDFs mostly because
we use the $\Delta \chi^2=1$ criterion rather than a dynamic tolerance
condition as in the MMHT analysis. 
We have also compared the profiled PDFs with alternative scale choices and
found those with NNLO predictions are less sensitive to the choice of
scale.

\section{Conclusion}
In conclusion we have presented details on calculation of next-to-next-to-leading
order QCD corrections to massive charged-current coefficient functions in
deep-inelastic scattering. 
We focus on the application to charm-quark production in neutrino scattering on
fixed target that can be measured via the dimuon final state as in the NuTeV
and CCFR experiments.
We construct a fast interface to the calculation so for any parton
distributions the dimuon cross sections can be evaluated within milliseconds
by using the pre-generated interpolation grids.
The NNLO predictions thus can be conveniently included in future global analyses
of QCD involving the dimuon data.
We further compare the dimuon data with the NNLO predictions using various PDFs and confirm
the pull of the ATLAS data on the strange-quark PDFs.
Moreover, we study impact of the NNLO corrections on the extraction of strange-quark
PDFs in the context of Hessian profiling, and find with the NNLO predictions
the dimuon data tend to favor larger strange-quark PDFs than with the
NLO predictions.
The fast interface together with the interpolation grids for dimuon
cross sections are publicly available upon request.
A definite conclusion on the potential inconsistency of the DIS and ATLAS
data awaits the ongoing global fits by the PDF analysis groups.  

\begin{acknowledgments}
JG would like to thank E. Berger, P. Nadolsky, R. Thorne, 
C.-P. Yuan and HX Zhu for useful conversations, and Southern Methodist University for
the use of the High Performance Computing facility ManeFrame.
The work of JG is sponsored by Shanghai Pujiang Program.
\end{acknowledgments}

\bibliography{strange}
\bibliographystyle{jhep}

\end{document}